\documentclass[aps, prd, twocolumn, superscriptaddress, nofootinbib]{revtex4}
\usepackage{graphicx}% Include figure files
\usepackage{dcolumn}% Align table columns on decimal point
\usepackage{amssymb}% outlined letters

\begin{document}

%Macros
\newcommand{\Eq}[1]{\mbox{Eq. (\ref{eqn:#1})}}
\newcommand{\Fig}[1]{\mbox{Fig. \ref{fig:#1}}}
\newcommand{\Sec}[1]{\mbox{Sec. \ref{sec:#1}}}

\newcommand{\PHI}{\phi}
\newcommand{\PhiN}{\Phi^{\mathrm{N}}}
\newcommand{\vect}[1]{\mathbf{#1}}
\newcommand{\Del}{\nabla}
\newcommand{\unit}[1]{\;\mathrm{#1}}
\newcommand{\x}{\vect{x}}
\newcommand{\ScS}{\scriptstyle}
\newcommand{\ScScS}{\scriptscriptstyle}
\newcommand{\xplus}[1]{\vect{x}\!\ScScS{+}\!\ScS\vect{#1}}
\newcommand{\xminus}[1]{\vect{x}\!\ScScS{-}\!\ScS\vect{#1}}
\newcommand{\diff}{\mathrm{d}}

\newcommand{\be}{\begin{equation}}
\newcommand{\ee}{\end{equation}}
\newcommand{\bea}{\begin{eqnarray}}
\newcommand{\eea}{\end{eqnarray}}
\newcommand{\vu}{{\mathbf u}}
\newcommand{\ve}{{\mathbf e}}

        \newcommand{\vU}{{\mathbf U}}
        \newcommand{\vN}{{\mathbf N}}
        \newcommand{\vB}{{\mathbf B}}
        \newcommand{\vF}{{\mathbf F}}
        \newcommand{\vD}{{\mathbf D}}
        \newcommand{\vg}{{\mathbf g}}
        \newcommand{\va}{{\mathbf a}}

%=====================================================================
%=====================================================================
%=====================================================================

\title{The case for testing MOND using LISA Pathfinder}

\newcommand{\addressImperial}{Theoretical Physics, Blackett Laboratory, Imperial College, London, SW7 2BZ, United Kingdom}

\author{Jo\~{a}o Magueijo}
%\email{magueijo@ic.ac.uk}
\affiliation{\addressImperial}

\author{Ali Mozaffari}
%\email{ali.mozaffari05@imperial.ac.uk}
\affiliation{\addressImperial}

\date{\today}

\begin{abstract}
We quantify the potential for testing MOdified
Newtonian Dynamics (MOND) with LISA Pathfinder (LPF), should a
saddle point flyby be incorporated into the mission. We forecast
the expected signal to noise ratio (SNR) for a variety of
instrument noise models and trajectories past the saddle. For
standard theoretical parameters the SNR reaches middle to high double
figures even with modest assumptions about instrument performance
and saddle approach. Obvious concerns, like systematics arising from LPF 
self-gravity, or the Newtonian background, are examined and 
shown not to be a problem. We also investigate the impact of a negative
observational result upon the free-function determining the theory.
We demonstrate that, if Newton's gravitational constant is constrained 
not be re-normalized by more than a few percent, only contrived 
MONDian free-functions would survive a negative result. There are
exceptions, e.g. free-functions not asymptoting to 1 in the Newtonian limit,
but rather diverging or asymptoting to zero (depending on their mother
relativistic MONDian theory). Finally, 
we scan the structure of all proposed relativistic MONDian theories,
and classify them with regards to their non-relativistic limit, finding
three broad cases (with a few sub-cases depending on the form of the free
function). It is appears that only the Einstein-Aether formulation, and 
the sub-cases where the free-function does not asymptote to 1 in other
theories, would survive a negative result without resorting to ``designer''
free-functions. 
\end{abstract}


\maketitle

%=====================================================================
%=====================================================================
%=====================================================================

\section{Introduction}
Einstein's theory of General Relativity (GR) and the $\Lambda$CDM
standard model are two cornerstones of modern cosmology. They
posit that the gravitational effects of large scale structures in
the universe (such as galaxies and clusters of galaxies) cannot be
explained by luminous, baryonic matter alone, but rather that an
additional cold (pressureless) and dark (non-luminous) matter
component is needed. However, in the absence of direct
observational evidence for dark matter, it remains nothing but a
useful calculational device. For as long as this is true, it is
scientifically healthy to explore alternative explanations for the
anomalous gravitational dynamics, namely by modifying the theory
of gravity itself.

MOdified Newtonian Dynamics (or MOND~\cite{Milgrom:1983ca}) is one
such scheme valid in the non-relativistic regime. It was first
proposed to explain observed dynamical properties of galaxies
without invoking dark matter. More recently it has been
incorporated into relativistic 
theories~\cite{BSTV,aether,aether1,Milgrom:2009gv,Milgrom:2010cd}, 
following from the ground breaking proposal of
``TeVeS'' by Bekenstein~\cite{teves}. Relativistic extensions are
needed for reasons beyond logical completeness: they are required
to explain, for example, phenomenology associated with lensing and
cosmology~\cite{kostasrev},  where dark matter
is also usually employed. When the whole
picture is assembled the conflict between MOND and dark matter
leaves considerable scope for doubt over the interpretation of new
astrophysical and cosmological data. A fair comparison requires
re-evaluating, within each approach, the whole set of assumptions
underlying the new observations. For this reason the debate would
benefit from a direct probe, in the form of a laboratory or Solar
System experiment. This has been proposed in various forms, namely
in planetary data~\cite{Blanchet} appealing to the exterior field
effect~\cite{Milgromss}.

The fact that MOND predicts anomalously strong tidal stresses in
the vicinity of saddle points of the Newtonian potential has been
advocated as one such  decisive direct
test~\cite{Bekenstein:2006fi}. The forthcoming LISA Pathfinder
mission~\cite{LISA} presents the perfect opportunity for its
realization, as a preliminary feasibility study has
demonstrated~\cite{Bevis:2009et,companion}. The purpose of this paper is to
provide a detailed quantitative evaluation of the power of a
MONDian saddle test using LPF, predicated upon a scenario where a
mission extension is granted. The extension would involve  
redirecting the spacecraft from Lagrange point L1 to a saddle of the
Earth-Moon-Sun system~\cite{companion} once its nominal mission at L1
is completed. In establishing the scientific case our efforts in
this paper are twofold.

In the first part of the paper 
we propose some basic data analysis tools and evaluate
their expected performance. These tools are an adaptation of the
``noise-matched filters'' employed in gravitational wave 
detection~\cite{sathya}. Their implementation 
benefits from a major simplification: for a saddle test we do know
the template's starting point in time. A number of pitfalls
and potential systematics found in detection of gravitational 
waves are therefore expected to be absent. The filter's optimal signal 
to noise ratio (SNR) allows us to quantify with a single number the
predicted outcome for any experiment. Assuming a ``standard'' MONDian
theory, the unknowns reduce to
the instrument performance (the noise properties) and the
trajectory past the saddle (its impact parameter). For
each of these we can condense the expected outcome of a
LPF test in a single number: the forecast SNR assuming MOND is
correct.

Our central results are in Section~\ref{secsnr}, particularly in
Figs.~\ref{fig:SNR contours} and~\ref{fig:SNR-improved}, 
where the optimal SNR is plotted
against noise level and saddle impact parameter. In a cataclysmic
scenario for instrument performance and saddle approach we'd still
achieved ${\rm SNR}\approx 5$. For less pessimistic assumptions, high
double figures are easily reached. We examine the effect of the
spacecraft speed as it flies past the saddle, showing that
just about any typical speed will turn out to be optimal. This is due to a
remarkable coincidence, spelled out in Section~\ref{systematics} and in
the concluding section of this paper. In Section~\ref{systematics}
we also show that possible systematic errors, such as
self-gravity or the Newtonian background, are in fact harmless.

In the second part of this paper, and complementing the work just described, 
we spell out the generality, or otherwise,  of the conclusions in the 
first part, and examine the implications of a negative observational result. 
Just how comprehensively would the failure to detect the predicted high SNRs
rule out the MONDian paradigm as a whole? 
As explained in Section~\ref{theory} the large
menagerie of proposed relativistic MONDian theories practically all 
reduce to the same non-relativistic limit as TeVeS, and virtually all 
theories fall 
into 3 categories. One then is left with a free-function, $\mu$, and
the question is, how 
much leeway does it provide for evading a negative result?
In Section~\ref{frefn} we review previously proposed free functions,
rewriting them under a unified notation. We then lay down conditions
for what should be permissible {\it simple}
free functions at the most basic level (by simple we mean with a minimum
number of regimes and scales). Briefly we require that: (I) The
theory shouldn't renormalize Newton's gravitational constant by
more than a few percent in the Newtonian regime; (II) The
theory should predict the usual MONDian effects when the Newtonian
acceleration drops below acceleration scale $a_0$. These
constraints were implemented in TeVeS and set the standard for a
viable theory with useful astronomical implications. We show that
{\it all natural functions satisfying these conditions result
in similar SNRs for an LPF saddle test, as long as the impact parameter
is smaller than $400\unit{km}$}. The only exceptions are free-functions
with a divergence (and with the rest of their domain excised) such as
those suggested in~\cite{Angus,FamaeyN}; however 
these {\it may} fall foul of existing Solar system constraints and other 
requirements~\cite{Fam-gaugh,FamaeyN,ssconst,Gentile}. We leave 
for a future publication a more detailed examination of these functions.

Of course one may open the doors to free-functions with more structure:
e.g. functions with three or more regimes instead of the minimal two.
In Section~\ref{nullres} we quantify how contrived the free function
$\mu$ would have to be, for the theory 
to survive a negative result. We find that, for bounded functions, 
only a $\mu$ turning from 1 (Newtonian regime) into an intermediate 
power-law, $\mu\propto z^n$, and only then into the MONDian $\mu\propto z$, 
would be viable. The intermediate $n$ would have to be very different 
from 1 even with undemanding requirements on impact parameter and noise.
Another possibility are the free-functions suggested
in~\cite{Zhao,Bruneton,McGaughMW,Zhaoeco}, for which galaxy rotation fits
and Solar system constraints are combined to motivate more structured
free-functions. All of these invoke three regimes (and so have two
implied scales). Even then, in Section~\ref{nullres1} we show that 
these would be within striking distance for a saddle test
with LPF.
Thus, with the exception of diverging $\mu$, 
only ``designer'' $\mu$ would bypass a negative result. 
Although this conclusion is derived for TeVeS-like theories 
it might be more general. 
With the usual honorable exceptions and provisos, 
the Einstein-Aether formulation~\cite{aether,aether1}
and a particular case of Milgrom's bimetric 
theory~\cite{Milgrom:2009gv,Milgrom:2010cd}
might be the only relativistic 
realizations of MOND to realistically survive a negative result
from a saddle test, a matter we'll examine in a future publication.

We conclude that a LPF test has both the power to detect
MOND with a high SNR should it be true, and to  rule it out,
should a negative result be obtained.

\section{MONDian theories}\label{theory}
One can find in the literature a large number of relativistic
MONDian theories. It is important to note that their complexity
and differences arise from the requirement that they should
explain relativistic phenomena (such as lensing and structure
formation) without appealing to dark matter. However, in the
non-relativistic regime, almost all of them reduce to the
non-relativistic limit of TeVeS, which will be the focus of this
paper. There are exceptions, however, and we spell them out here.
In general the large profusion of relativistic MONDian theories
reduces to only 3 types of non-relativistic limits, which we'll
label type I, II and III.

\begin{itemize}
\item {\bf Type I} In these theories the non-relativistic dynamics
results from the joint action of the usual Newtonian potential
$\Phi_N$ (derived from the metric via $g_{00}\approx
-(1+2\Phi_N)$) and a ``fifth force'' field, $\phi$, responsible
for MONDian effects. The total potential acting on
non-relativistic particles is their sum: \be \Phi=\Phi_N+\phi\; .
\ee 
Whilst the Newtonian potential satisfies the usual Poisson
equation: \be \nabla^2 \Phi_N=4\pi G \rho \ee the field $\phi$ is
ruled by a non-linear Poisson equation: \be \nabla \cdot
\left(\mu(z)\nabla \phi\right) = \kappa G \rho, \ee where, for
convenience, we pick the argument of the free function function
$\mu$ as: \be
z=\frac{\kappa}{4\pi}\frac{\vert\nabla\phi\vert}{a_0} \ee where
$\kappa$ is a dimensionless constant and $a_0$ is the usual MOND
acceleration. In Section~\ref{frefn} we will say more on admissible
functions $\mu$, but in general we require that $\mu\rightarrow 1$
when $z\gg 1$ and $\mu\sim z$ for $z\ll 1$. (We use letter $z$
instead of $y$ to prevent a common source of confusion in the
literature; see Section~\ref{notation}.)

\item {\bf Type II} In these theories we also have $\Phi=\Phi_N
+\phi$, but now the field $\phi$ is ruled by a driven linear
Poisson equation, whose source depends on the Newtonian potential.
In order to facilitate comparison with Type I theories (as
explained below) we write the equation for $\phi$ in these
theories as: \be \nabla^2 \phi =
\frac{\kappa}{4\pi}\nabla\cdot\left(\nu(v)\nabla \Phi_N \right)
\ee where the argument of free function $\nu$ is given by \be v=
\left(\frac{\kappa}{4\pi}\right)^2\frac{\vert\nabla\Phi_N
\vert}{a_0}  \ee and we require that $\nu\rightarrow 1$ when $v\gg
1$ and $\nu\sim 1/\sqrt{v}$ for $v\ll 1$. However, it is possible
in some models that
that $\nu\rightarrow 0$ in the same limit, with qualitatively
very different implications. To distinguish these two cases we
call the latter ($\nu\rightarrow 0$) type IIA theories and the former 
($\nu\rightarrow 1$) type IIB theories.

\item {\bf Type III} This was the original non-relativistic
MONDian proposal, derived from a non-relativistic action principle
(the so-called AQUAL~\cite{aqual}). Crucially, here non-relativistic particles
are sensitive to a single field $\Phi$ which satisfies a
non-linear Poisson equation: \be \label{typeiii}\nabla\cdot\left(\tilde{\mu}(x)
\nabla \Phi \right) = 4 \pi G \rho\; . \ee Again, $\tilde\mu$ is a
free function with a suitably chosen argument: \be x=
\frac{\vert\nabla\Phi\vert}{a_0}\ee so that $\tilde \mu\rightarrow
1$ when $x\gg 1$ and $\tilde \mu\sim x$ for $x\ll 1$.
\end{itemize}

A couple of remarks  on this classification are in order. 
Firstly, note that in some theories 
$\Phi=\Xi\Phi_N+\phi$, where $\Xi\approx 1$ is a factor evolving slowly on
cosmological time scales. We have set $\Xi=1$ throughout, for simplicity.
Secondly, it is obvious that the two equations ruling type II theories may 
be rewritten as a single equation, ruled by a redefined $\nu$. This is
a cosmetic matter and is beside the point. The real matter is whether 
$G$ is renormalized, with regards to the bare $G$ and the 
cosmological one (a matter we will presently discuss in detail). This is 
encoded in the limiting behaviour of the free function ruling the 
non-relativistic equations, whether they're written as a single
field or two fields system. 
It has been argued that in some relativistic formulations
of type II theories the bare $G$, that appearing in cosmology and
the total $G$ ruling the non-relativistic equation are the same. We call 
such theories type IIA, and for them $\nu\rightarrow 0$ in the Newtonian
limit. Otherwise we call then type IIB,  with a $G$ 
renormalization, and with $\nu\rightarrow 1$. 
As we will see the matter has crucial implications.

Virtually all relativistic MONDian theories proposed in the
literature fall into these categories. TeVeS,  the pioneering
relativistic MONDian theory~\cite{teves}, has type I limit, but Sanders'
stratified theory~\cite{BSTV} is also type I. Milgrom's bimetric 
theory~\cite{Milgrom:2009gv,Milgrom:2010cd} can be
either type I or type II, depending on details. Einstein-Aether
theories~\cite{aether,aether1} are unique in that they have a
 non-relativistic limit of
type III. Often authors have attended to different
considerations and constraints, so the parameter $\kappa$ has been
taken to be different. However, as we'll point out below, had the
same considerations been employed, the value of $\kappa$ would
have to be comparable. 

There are significant differences between the non-relativistic
limits listed above. The most radical distinction bundles together
type I and IIB theories in opposition to type IIA theories and 
the single relativistic
theory leading to a type III limit. 
Because in type I and IIB
theories non-relativistic particles are sensitive to two fields,
which mimic each other in the Newtonian regime, 
the gravitational constant is effectively renormalized. In the 
Newtonian regime (non-relativistic limit, high total Newtonian
force), we have $\mu\approx 1$ or $\nu\approx 1$, and so $\phi$
becomes proportional to the Newtonian potential: \be\label{rat} \phi\approx
\frac{\kappa}{4\pi}\Phi_N\; , \ee i.e. $\phi$ doesn't vanish but
rather shadows $\Phi_N$ multiplied by $\kappa/4\pi$. This
``renormalizes'' the gravitational constant: \be G_{Ren}\approx
G{\left(1+\frac{\kappa}{4\pi}\right)}\; , \ee and $G_{Ren}$ is the
gravitational constant measured, say, by the Cavendish experiment.
%(affected by both the Newtonian potential and a fifth force from
%which it is undistinguishable).
Nevertheless cosmology (for example, 
Friedmann's equations) is sensitive  to the ``bare'' $G$.
Constraints arising from Big Bang nucleosynthesis therefore
require $\kappa$ to be of the order of $\sim 0.01$ or smaller.
Structure formation considerations may fix further the value of
$\kappa$ (see~\cite{kostasrev} and references therein). 
The conclusion is that in the non-relativistic regime
the field $\phi$ must be suppressed when $a_N=|\nabla\Phi_N|$ 
is very large.

\begin{figure}\begin{center}
\resizebox{1.\columnwidth}{!}{\includegraphics{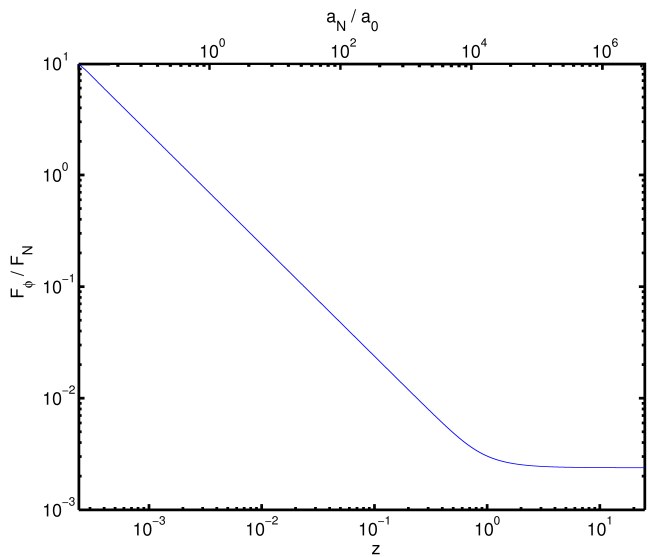}}
\caption{\label{fig:Fphi/FN plot analytical mu}{Log plot of ratio
between the MONDian and Newtonian forces, $F_\phi/F_N$, against $z=(k
/ 4\pi) |F_\phi| / a_0$ (bottom axis) and $F_N/a_0$
(top axis).  So that $F_N\sim F_\phi$ when 
$F_\phi\sim a_0$ (and so $z=\kappa/4\pi$; also $F_N\sim a_0$) 
and at the same time
have $F_\phi/F_N\sim
\kappa  /4\pi\ll 1$ in the Newtonian regime ($z\gg 1$, $F_N\rightarrow\infty$),
we must trigger MONDian
behaviour in $\phi$ at accelerations much larger than $a_0$
(when $z\sim 1$).
}} \end{center}
\end{figure}

However, astrophysical applications of type I and IIB 
theories require that when $a_N<a_0$ the {\it total} $\Phi$ must have MONDian
behaviour. This requires {\it simultaneously} that $\phi$ be in the MONDian
regime and that $\phi$ be the dominant contribution.
But this means that we must switch on MONDian behavior in $\phi$ at
Newtonian accelerations $a_N$ much higher than $a_0$. Only thus
may the relative importance of $\phi$ start increasing with
decreasing $a_N$ so that when $a_N$ drops below $a_0$ it has
caught up with $\Phi_N$. 
Assuming the free function turns from 1 to a single power-law (and
ignoring the MONDian magnetic field where appropriate) we 
have $F_\phi/F_N\propto 1/\sqrt{F_N}$ once MONDian behavior in
$\phi$ has been triggered. Given (\ref{rat}) we should therefore
trigger MONDian behavior in $\phi$ for: \be a_N<a^{trig}_N\approx
{\left(\frac{4 \pi}{\kappa}\right)}^2a_0\;, \ee (with
$a^{trig}_N\sim 10^{-5} \unit{m s}^{-2}$ for typical $\kappa$) or
equivalently \be \vert\nabla \phi\vert
<a_\phi^{trig}=\frac{4\pi}{\kappa} a_0 \ee also much larger than
$a_0$. This point is illustrated in 
Fig.~\ref{fig:Fphi/FN plot analytical mu}.

This simple argument fails should $\mu$ be divergent (with part of its
domain excised). Then, $F_\phi$ goes to a constant as $F_N\rightarrow\infty$, 
and so $G_{Ren}=G$ (see ~\cite{Angus,FamaeyN}). 
Consequently, it is possible to have $a_N^{trig}\approx
a_0$, without fine-tuning the free-function or inducing 
unduly different $G_{Ren}$, in these theories. 
Such functions, however, {\it may} 
have other problems, rendering them non-viable.
(This is explained further in in Sections~\ref{frefn} and~\ref{nullres1};
see Fig.~\ref{fig:Fphi/FNnew},  which contains the
counterpart of Fig~\ref{fig:Fphi/FN plot analytical mu}
for these theories.) For the same reason the simple argument 
laid down in the previous two paragraphs also fails
for type IIA theories, for which $G$ is not renormalized.

Excluding the very last proviso, our
considerations apply equally to type I and IIB
theories. We have parametrized the free-function $\nu$ for type
IIB theories in a way (at odds with the literature) which allows
comparison of ``like with like'' with type I
theories. Thus, for the same $\kappa$ both types of theory
renormalize the gravitational constant by the same amount. They
also then predict $\phi\sim \Phi_N$ for $a_N\sim a_0$ when
their functions $\mu$ or $\nu$ trigger MONDian behavior at $z\sim
1$ and $v\sim 1$, for type I and IIB theories, respectively.

The features we have highlighted explain the large size of the bubbles around
the saddle inside which type I and IIB theories display anomalously
large tidal stresses. These bubbles are large (of the order
$r_0\approx 380\unit{km}$ for the Earth-Sun saddle) because they
represent the region where the field $\phi$ has started to behave in a
MONDian fashion. This is given by the region where
$a_N<a_N^{trig}$ and {\it not} where $a_N<a_0$, as might be
naively expected (and indeed it can be easily computed that 
$a_N\sim a_N^{trig}$ around $r_0$). 
It is important to stress that in a LPF saddle
test we are probing the regime where $\phi$ has gone fully MONDian
but hasn't yet dominated $\Phi_N$, something that happens at total
Newtonian forces in the range $a_0<a_N<a_N^{trig}$.
In spite of the dominance of $\Phi_N$ the MONDian signal in $\phi$
can be detected  because, as we shall see,
it has a distinctive spatial variation,
whereas the Newtonian tidal stress is just a DC component. The
experiment is sensitive to the time Fourier transform of the
signal with a sensitivity that peaks at the MOND frequency (and is
very poor for a DC component due to $1/f$ noise). 
In contrast, in type IIA and III theories MONDian effects are only triggered
for $a_N\sim a_0\sim 10^{-10}\unit{m}\unit{s}^{-2}$, resulting in
very small bubbles (a few meters across).

Even though these considerations place type I and IIB theories 
in the same basket with regards to a saddle test, 
they have a significant difference. A well known
technicality is that type I and III theories have a curl term
(sometimes dubbed a ``magnetic field''), in the sense that if one attempts
to linearize their non-linear Poisson equation by introducing an auxiliary
vector field (e.g. $\mu\nabla\phi$ for type I theories) this field has
non-zero curl. The same doesn't happen for type II theories, which are
already linear in $\phi$ (but driven by a function of the Newtonian 
field, $\nu\nabla \Phi_N$, which has a curl). This turns out to
have a significant quantitative (but not qualitative) 
effect upon the magnitude of the saddle tidal stresses.
The magnetic field is known to soften the anomalous tidal
stresses around the saddle points in type I theories, as explained 
in~\cite{Bekenstein:2006fi}.  As we'll show in~\cite{aliqmond},
type IIB theories have a quantitatively stronger saddle signal than type I 
(once their $\kappa$ are adjusted to produce the same physical properties),
due to the absence of this curl field\footnote{Note that if one defines
the curl field as $\nabla\phi - \nu \nabla\Phi_N$ in type II
theories (in analogy to $\mu \nabla\phi -\nabla\Phi_N$ in type I
theories), then these theories do have a curl field.}.

In this paper we focus on Type I MONDian theories, but in
the conclusions explain why our results are qualitatively
applicable to Type IIB theories too (indeed the SNR forecast here
are higher for type IIB theories~\cite{aliqmond}). Type IIA and 
III theories are the only ones to fall through the LISA Pathfinder net.

\section{The Signal to Noise expected from a saddle
flyby} \label{secsnr}
\begin{figure}
\resizebox{\columnwidth}{!}{\includegraphics{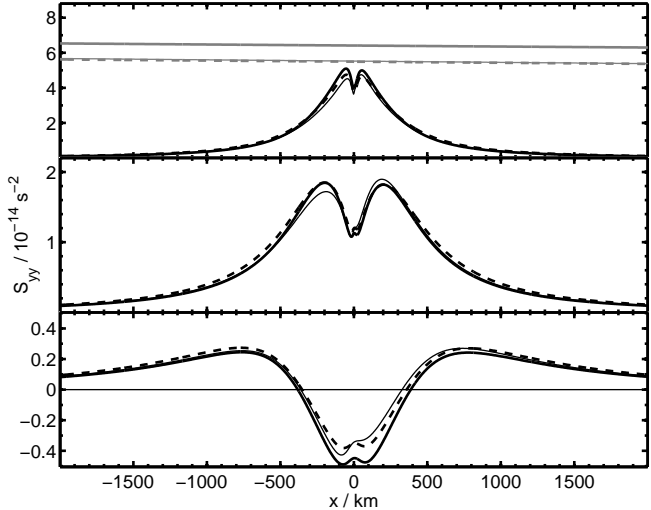}}
\caption{\label{fig:3bodyStress}The transverse MOND stress signal
$S_{yy}$ along the lines $y=25$, $100$ and $400 \unit{km}$ (top to
bottom), for the Sun-Earth saddle, taking the effect of the Moon
into account. The different lines represent different lunar
phases: new Moon (thick, black, solid), full Moon (thick, black,
dashed) and with the Moon appears $18^\circ$ away from the Sun
towards positive $y$ (thin, black, solid). We also show, for the
$y=25\unit{km}$ case, the Newtonian stresses (grey) rescaled by
$\kappa/4\pi$ (see text for discussion).}
\end{figure}

The quantitative predictions for type I theories have been
extensively studied using both analytical methods resorting to 
simplifying assumptions~\cite{Bekenstein:2006fi} and  
numerical techniques~\cite{Bevis:2009et}
including all the complications of the problem,
such as the perturbing effect of the Moon and planets. 
Figure~\ref{fig:3bodyStress} has been borrowed from~\cite{Bevis:2009et} to
illustrate the expected tidal stress along lines missing the saddle at
25, 100 and 400 km. As in~\cite{Bevis:2009et}, we adopted 
a coordinate system with $x$ aligned along the
Sun-Earth axis and centered at the saddle and considered 
trajectories parallel to $x$ ($y=b$ lines, where $b$ is the
impact parameter), but other trajectories are easy to implement.
Due to a number of practical issues~\cite{companion}, 
only transverse tidal stresses can be measured, 
say the $S_{yy}$ component.

Predictions are cast in the form of tidal stresses because 
this is what is directly measured by the instrument. LPF measures  
the relative distance between the masses, but this is converted 
into a relative acceleration (or its Fourier transform in time). 
Up to a factor dependent on the proof mass separation, the
measurement is therefore one of tidal stress along the direction
linking the two masses (with further masses, other tidal stress
components would become accessible). In line with this statement,
noise evaluations and forecasts are expressed in terms of tidal stress
or relative accelerations; one should use the inter-mass separation
to convert between the two.

It is of paramount importance to note that the 
field $\phi$ produces both a MONDian
effect and a Newtonian pattern, associated with a rescaling of $G$
in the Newtonian limit. The properly MONDian stress is
therefore: \be\label{Sij} S_{ij} = -\frac{\partial^2
\phi}{\partial x_i
\partial x_j} + \frac{\kappa}{4 \pi} \frac{\partial^2
\Phi^N}{\partial x_i \partial x_j}  \label{S
ij defn}\ee 
i.e. we must subtract from $\phi$ its component included in the Newtonian
background, which is $\Phi_N$ rescaled by $\kappa/4\pi$
(see~\cite{Bevis:2009et} for more details). In performing this exercise
it is essential that $\phi$ and $\Phi^N$ have been
solved to the same degree of accuracy. In Section~\ref{newt} 
we will discuss the impact of an imperfect subtraction of the Newtonian 
component. 

The data analysis task in hand is therefore to detect a ``wave form'' of 
this type with the instrument aboard LPF. As a first hack at the problem, we
evaluate the performance of noise matched filters. Matched
filtering is a well-known data analysis technique used for
efficiently digging a signal with a known shape out of  noisy
data~\cite{Helstrom,sathya}. The technique is extensively used in
the search for gravitational waves. The idea is to correlate a
time series $x(t)$ with an optimized template designed to provide
maximal signal to noise ratio (SNR), given the signal shape $h(t)$
and the noise properties of the instrument. Generally we have
$x(t)=h(t-t_a)+n(t)$, where $t_a$ is the signal ``arrival time''
and $n(t)$ is a noise realization. We want to correlate $x(t)$ and
an optimal template $q(t)$, yet to be defined, according to: \be
c(\tau)=\int_{-\infty}^\infty x(t) q(t+\tau)dt\; , \ee where
$\tau$ ia a lag parameter, giving us essential leverage if we
don't know $t_a$ a priori. The average of $c$ over noise realizations is
the expected signal, $S$, and its variance is the square of the 
noise in the correlator,
$N^2$; the forecast signal to noise ratio is therefore $\rho=S/N$.
A straightforward calculation (under general
assumptions, namely the Gaussianity of the noise---more on this
later) shows that $\rho$ is maximized by choosing a template
with Fourier transform: \be \label{opttemp}{\tilde
q}(f)=\int_{-\infty}^\infty q(t) e^{2\pi i ft} dt= \frac{{\tilde
h}(f) e^{2\pi i f (\tau - t_a)}} {S_h(f)} \ee and setting the lag
$\tau$ to the arrival time, $\tau=t_a$. Here $S_h(f)$ is the power
spectral density (PSD) of the noise, conventionally defined from
\be{\langle {\tilde n}(f){\tilde n}^\star (f')\rangle}=\frac{1}{2}
S_h(f) \delta(f-f')\;,\ee (the factor of $1/2$ hails from the
tradition of taking one-sided Fourier transforms of the noise
auto-correlation----i.e. with $f>0$ only).
%Power spectral density of the noise Compute the noise
%auto-correlation. For a stationary process this is
%$\kappa(|t_1-t_2|)={\langle n(t_1) n(t_2)\rangle}$. Take the
%one-sided Fourier transform (for the one-sided PSD): \be
%S_h(f)=\frac{1}{2}\int_{-\infty}^\infty \kappa (\tau) \ee and this
%is the noise PSD (its square root is the ASD or amplitude spectra
%density CHECK).
The maximal SNR, realized by the optimal template, is then: \be
\label{SNR}\rho=\rho_{\mathrm{opt}} = 2 \left[ \int_{0}^\infty \,
df\frac{ \left |\tilde h(f)\right|^2 }{S_h(f)} \right]^{1/2}\; .
\label{eq:snr}\ee Notice that the optimal template, $q(t)$,
defined by (\ref{opttemp}) is not the signal, $h(t)$, but rather a
filtered version with a pass where the noise is low and a cut
where the noise in high. Also, the optimal SNR given by
(\ref{SNR})  is not the energy in the signal but an integrated
signal power weighted down by the noise PSD.

These techniques are run of the mill in gravitational wave
detection, where the arrival time of a signal is often not
known\footnote{There are exceptions, for example if the signal
comes from a supernova or any other source for which there is an
extrinsic method, typically in the optical domain, for flagging
the source of gravity waves.}. For example, for a chirping signal,
even if we have a fair idea of the shape of the signal, we can't
know when a binary coalescence is to take place. We therefore have
to shift the template Fourier transforms, ${\tilde h}(f)$, by all
possible phases, until the maximal SNR is obtained, should there
be indeed a signal. This adds an extra parameter to the fit and
may also be the source of spurious detections. It affects the management
of $1/f$ noise and increases the false alarm rates (as effectively 
we have a number of trials equal to the total observation time 
divided by the duration of the template). 
This problem is absent in the context of our test, where $t_a$ is known
since we do know where the saddle is and therefore where the
signal is meant to start in the time-ordered series. A natural truncation
in integration time $T$ is also present, simplifying $1/f$ dealings.

It has been estimated\footnote{These uncertainties are of a practical nature
and should not be confused with theoretical uncertainties. It can be 
estimated that the MOND saddle cannot be shifted with respect to the 
Newtonian saddle by more than a meter, and this is just an upper bound.} 
that the saddle can be pin pointed to about
a kilometer and the spacecraft location determined to about
10 km even with most basic tracking methods\footnote{S. Kemble, 
private communication; to be published.}. The effect this has on 
SNRs is negligible, indeed the SNR grids we are about to
present have this sort of resolution. Thus, we can
simply set $t_a=0$ with an appropriate choice of conventions and
set to zero the time lag $\tau$ in the correlator $c$, to
achieve optimal results. For all practical purposes the starting time
is indeed known.  To the same degree of approximation, we also know the
spacecraft trajectory and velocity with respect to the saddle.

Given a spacecraft trajectory, the conversion of tidal stresses
(such as those depicted in Fig.~\ref{fig:3bodyStress}) 
into a template in time, $h(t)$,
is then trivial. For a setup such as the one described
above we have $h(t)=S_{yy}(vt,b,0)$, where $v$ is the
velocity of the spacecraft, and 
$t=0$ corresponds to the point of closest saddle approach. In a more
general setup, for an approximately constant velocity $\bf v$,
a closest approach vector $\bf b$, and with the masses aligned along
unit vector $\bf n$, we have \be \label{hoft}
h(t)=n^i n^j S_{ij}({\bf b} +
{\bf v}t ) \; .\ee This template should be Fourier transformed and, given a
noise model, used to produce an optimal template, using a noise
matched filter. Its SNR can then be evaluated.

\begin{figure}\begin{center} {\includegraphics[scale=0.4]{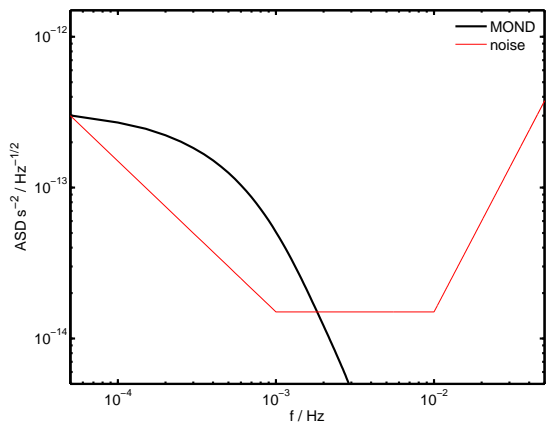}}\\
\caption{\label{fig:signal + noise ASD} The amplitude spectral
density (ASD) of the MOND tidal stress signal for a trajectory
with $b=50\;\unit{km}$ and $v=1.5\unit{km}\unit{s}^{-1}$,
compared to the ASD of the basic noise model described in the text, 
assuming a baseline
of $1.5\times10^{-14}\unit{s^{-2}/\sqrt{Hz}}$. This scenario generates
a SNR of 28.}
\end{center}\end{figure}

\begin{figure}\begin{center}
\resizebox{1.\columnwidth}{!}
%{\includegraphics{SNRcontour.eps}}
{\includegraphics{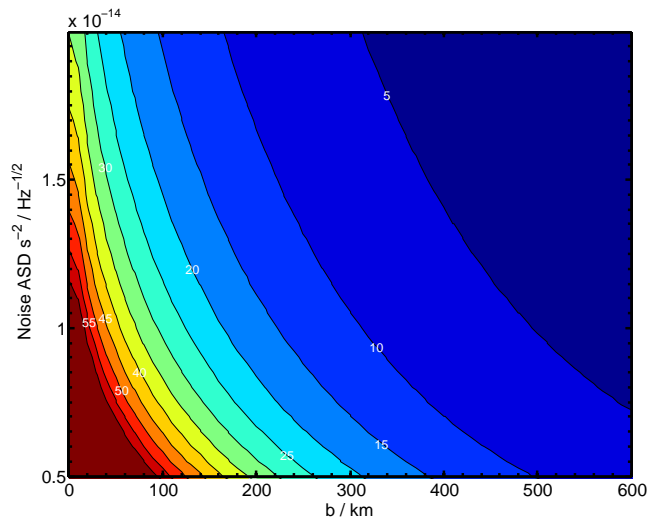}}
\caption{\label{fig:SNR contours}  Signal to Noise ratio
contours, for various impact parameters up to 600km and base noise
ASD. We set the spacecraft velocity at 1.5km s$^{-1}$. 
Calamitous assumptions would still lead to SNR of 5. More optimistic
ones ($b$ around 50km, noise half way up the scale) would lead to
SNRs easily around 50.}\end{center}
\end{figure}

To gain some intuition on the nature of the signal  in
Figure~\ref{fig:signal + noise ASD} we plot the 
amplitude spectral density (ASD) of the signal, which is the
square root of the PSD:
\be P(f) = \frac{2}{T} \left| \int_{-T/2}^{+T/2} \diff
t \; h(t) \; e^{-2\pi i f t} \right|^{2} \ee where $f$ is the
frequency, $t$ is the time and $T$ is the integration period
(here taken conservatively to be $T= 2\times10^4\unit{s}$).
This can be directly compared to the noise 
ASD, the form usually quoted by experimentalists.  As a 
simplified LPF noise model (see~\cite{companion}), we 
assume that the noise is white
in the frequency range between $1$ and $10\unit{mHz}$,
i.e. we assume a constant baseline with ASD around
$1.5\times10^{-14}\unit{s^{-2}/\sqrt{Hz}}$. For lower frequencies we assume
$1/f$ noise and for higher frequencies that the noise degrades 
as $f^{2}$.  With these assumptions the noise and signal ASDs are 
plotted in Figure~\ref{fig:signal + noise ASD}, for typical parameters. 
As we can see, there's 
signal to noise of order 10 over a couple of decades, making it not 
surprising that the integrated SNR is in double figures 
(in this case around 28).

We can now run through the parameter space of the experiment and evaluate
SNRs. For example, let's assume $v=1.5\unit{km} \unit{s}^{-1}$ 
and explore impact
parameters up to 600 km. Let's also consider the effect of changing
the base line ASD of our noise model. The result is plotted
in Figure~\ref{fig:SNR contours}. We see that we'd need to miss the saddle
by more than 300 km to enter single figures in SNR, with typical noise
levels. For $b$ around 50 km 
a SNR of the order 30-40 is not unrealistic. Recent work has placed a
figure on the impact parameter around $b=10-50$~km within reach. In combination
with the expectations for the noise, this makes the test very promising
indeed. However we should now look at this preliminary analysis in 
more detail.

\section{Further discussion}\label{systematics}
In this Section we refine and discuss further the basic results 
presented in the previous Section. There is considerable uncertainty
regarding the details of the 
flyby trajectory, namely its speed. In Section~\ref{flyby}
we show that the effect of the speed is minimal, within
the range of speeds expected from any trajectory in the Moon-Earth 
system. In Section~\ref{noiseimprove} we present improved, more realistic
noise models, repeating the analysis with a best and worst
case scenario for instrument performance as understood 
at the time of writing. We
also outline work in progress, improving on noise matched filters and
on estimates of false alarm rates. Finally in Section~\ref{newt}
we discuss issues related to the background tidal stresses, namely the 
the Newtonian background and the spacecraft self-gravity.

\subsection{The impact of the spacecraft velocity}\label{flyby}
\begin{figure}\begin{center}
\resizebox{1.\columnwidth}{!}{\includegraphics{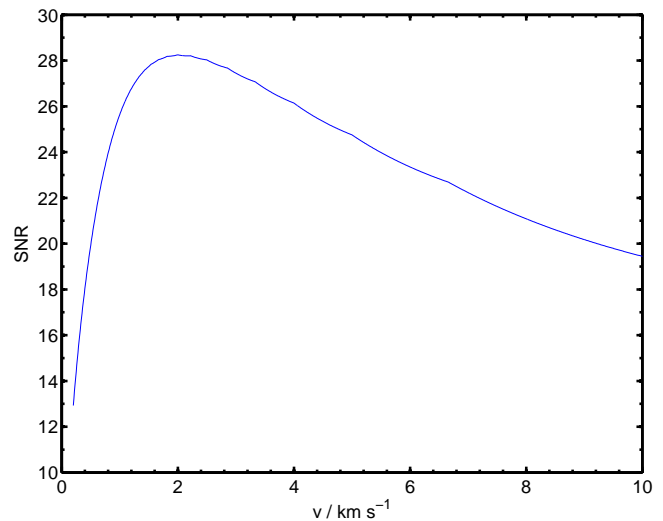}}
\caption{\label{fig:SNR velocity plot}Plot of SNR against
satellite velocity for an impact parameter of 50 km and a baseline
noise of $1.5\times10^{-14}\unit{s^{-2}/\sqrt{Hz}}$. We note a broad 
peak around $v = 2 \unit{km}\unit{s}^{-1}$. Higher speeds shift the signal to
higher temporal frequencies; however the rough speeds of all 
trajectories in the Earth-Moon system are already optimal, given
the noise properties of the instrument.}
\end{center} \end{figure}

What is the effect of the spacecraft velocity on the SNRs presented
in the previous Section? The question is relevant as it can
assist the strategy in designing flyby trajectories. 
Here we show that in practice all that matters is the 
trajectory location (impact parameter and possibly angle).
Within the range of realistic 
speeds, the SNRs do not vary substantially. The good news is
that due to a remarkable coincidence, these speeds are already
near optimal.

As Eq.~(\ref{hoft}) shows, the spacecraft velocity is the conversion factor
between the spatial scale of the tidal stress and the time scale
at which the instrument measures them. Of course, {\it in detail}, 
this has an effect on expected SNRs. Higher/lower 
speeds mean a faster/slower scanning of these spatial features, and
thus a shift of the template ${\tilde h}(f)$ to higher/lower frequencies, 
whilst keeping the noise ASD fixed. Therefore the SNR has to change. 
This is shown in Fig.~\ref{fig:SNR velocity plot}, for $b=50$~km
and a baseline noise of $1.5\times10^{-14}\unit{s^{-2}/\sqrt{Hz}}$.

We see that the SNR has a peak at $v = 2 \unit{km}\unit{s}^{-1}$. 
However  
this peak is very broad with respect to the type of variations that might be
expected from different trajectories leading from L1 to the 
saddle~\cite{companion}. For the rough range 
$v = 1.5-2.5 \unit{km}\unit{s}^{-1}$ the SNR varies in the range
27-28, approximately. For $v = 1-3 \unit{km}\unit{s}^{-1}$
(which is pushing it, in terms of real orbits)
the variations would be in the approximate range 25-28. The message is 
clear: get as close to the saddle as possible, never mind the speed. 
The speed will never be very far off the optimal.

This result can be understood qualitatively. As a crude estimate, 
anything moving in the Earth-Moon system  has a typical speed
of the order of 1~km~s$^{-1}$. The MONDian tidal stress for
the Earth-Sun saddle displays variations on a length
scale of the order of $100\unit{km}$. Therefore the MONDian signal will always 
be felt by LPF 
on a time scale of minutes, i.e. in the mHz range.
% in frequency space. 
This is just where the instrument noise is lowest, a remarkable
coincidence considering that the instrument was built to these specifications
for entirely different reasons (astrophysically motivated gravitational wave
templates have these time scales).
And yet the typical speeds and length scales of the problem
combine to make the instrument already optimal for a MONDian
saddle test.

\subsection{Improved noise models}\label{noiseimprove}
\begin{figure}\begin{center}
\resizebox{1.\columnwidth}{!}
%{\includegraphics{noisemodels.eps}}
{\includegraphics{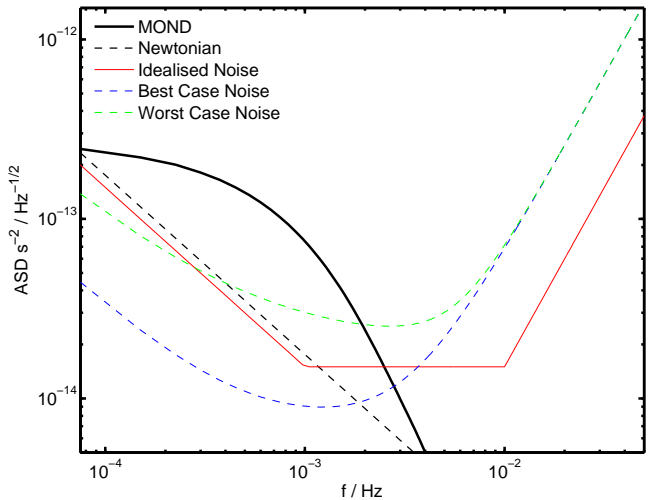}}
\caption{\label{fig:realnoise}This figure replots 
Fig.~\ref{fig:signal + noise ASD}, adding on the best and worst case
scenarios for more realistic noise models, as at the time of writing.
We have assumed a trajectory with the geometry described in the main 
text, with impact parameter of $b=50\;\unit{km}$ and 
velocity  $v=1.5\unit{km}\unit{s}^{-1}$.
We have also plotted the contribution of $\phi$ to the Newtonian background. } 
\end{center}
\end{figure}

A number of improvements to the noise model used in the previous
Section are possible. These are the subject of a paper in 
preparation~\cite{trenkprep}. Obviously there isn't 
a frequency region with white noise.
Instead, the noise is likely to be higher than modeled in 
Section~\ref{secsnr} at high frequencies 
but lower than expected at low frequencies. The turnover between
the two regimes is smooth, as depicted in Figure~\ref{fig:realnoise},
where we superimposed the simplified noise model used 
in Section~\ref{secsnr} with the more realistic
estimates for ASD for a best and worst case scenario.
It has been argued that the worst case scenario might be too
pessimistic and the best case scenario too optimistic, 
so we should take these two models as extremes. 

\begin{figure}\begin{center}
\resizebox{1.\columnwidth}{!}
{\includegraphics{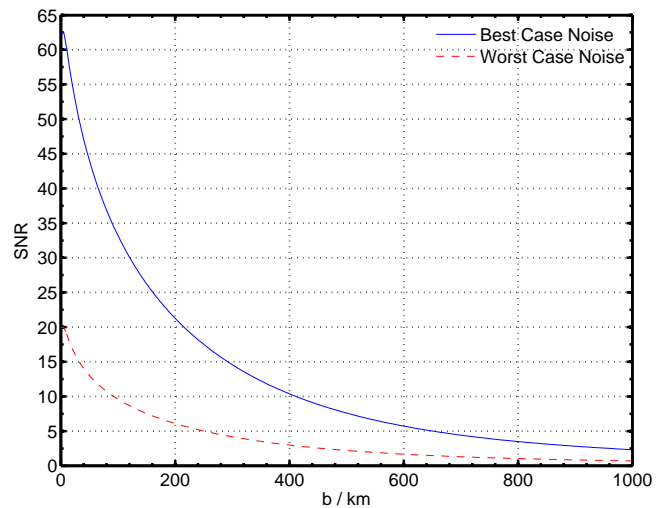}}
\caption{\label{fig:SNR-improved} The SNR for the improved noise models
(best and worst case scenario) assuming $v=1.5\unit{km}\unit{s}^{-1}$
for a variety of impact parameters $b$.
}\end{center}
\end{figure}

In Figure~\ref{fig:SNR-improved} we plot the SNR as a function of
impact parameter with $v=1.5\unit{km}\unit{s}^{-1}$, 
assuming the two extreme scenarios. As we can see, 
in the best case scenario we'd need
to miss the saddle by more than $650\unit{km}$ for the SNR to drop below 5. 
In the worst case noise scenario, however, that figure would shrink 
to about $250\unit{km}$.
For the currently expected $b\sim 50\unit{km}$ the SNR would be in the 
range $13-44$. In spite of the uncertainties, all scenarios lead to
optimistic prospects (and even overkill) regarding a detection. 
We stress that we will know what the noise is, {\it in situ}
and while on L1. Our forecasts are useful, but we should 
highlight that they'd become concrete, fixed numbers once the 
mission goes ahead.

We should add that even if the 
noise ASD is known, further issues complicate the simple data analysis 
procedures presented in the previous Section. Most notably 
the real noise is non-Gaussian and 
non-stationary. This {\it may} increase the probability of a ``false
alarm'', to use the jargon of gravitational wave detection. 
Putting a realistic figure to the probability of a false detection
requires having the instrument switched on before and after a
saddle flyby, characterizing the noise {\it in situ}, and evaluating
the false alarm rates with real noise. No prior modeling
can be a substitute for this. Nonetheless more realistic simulations of the 
instrument response and noise are possible. We are currently working on these.

The issue of false alarm rates is obviously central should there be
detection. But even just planning the experiment, it raises important
questions, e.g.: given these rates, is it better to sacrifice 
$b$ at the expense of multiple flybys, 
or should we put all our efforts into a single flyby with a $b$ as low 
as possible? Should the noise be approximately Gaussian and stationary, 
the probability of a false detection is simply~\cite{sathya}: 
\be
{\cal F}=N{\rm erfc}(\rho)
\ee
where $\rho$ is the optimal SNR, and $N$ is the number of trials.
In gravitational wave detection $N=O/T$, where $O$ is the total observation
time and $T$ the useful duration of the filtered template. The factor
$N$ can be
very large, so that even substantial SNRs (say 8 or 9) can produce
non-negligible rates ${\cal F}$. In gravitational wave detection
this nuisance can be mitigated by coincident observations. 
{\it We stress that no such problem is 
present here.} We do know where the saddle is
for all practical purposes, so $N=1$, removing the extra
factor enhancing the false alarm rate.

The high SNRs we've obtained at low $b$ suggest that it would not be 
advisable to sacrifice $b$ for the sake of multiple flybys, 
in order to reduce false alarm rate. This statement should be further
scrutinized using real noise. But even if it's true there is an important
sociological element. The reliability of any scientific claim rests on 
reproducibility. Should there be a positive detection, more than one flyby 
would go some way towards establishing the case for reproducibility.

\subsection{The Newtonian background and self-gravity}\label{newt}
\begin{figure}\begin{center}
\resizebox{1.\columnwidth}{!}{\includegraphics{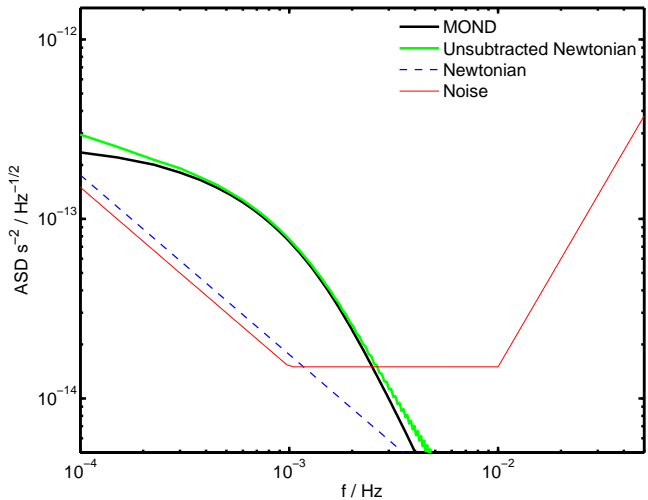}}
\caption{\label{fig:Tidal stresses ST1} ASD plot of the MONDian and
Newtonian signal (multiplied by $\kappa/4\pi$), as compared to the 
noise ASD. We consider the effect of subtracting the Newtonian component
in $\phi$. This only affects very small and very large frequencies.
}\end{center}
\end{figure}

We finish this section by examining two possible systematics that could 
plague a saddle test: the Newtonian background and the spacecraft self-gravity. 
These are natural concerns, but their impact is negligible.
In establishing this fact it is important not to confuse force and tidal 
stress. It is also essential to examine the Fourier components of the 
stress signal and distinguish a DC component from a signal peaking at 
frequencies to which the experiment is sensitive.  

The MONDian saddle signal has a spatial scale 
$r_0\approx 380\unit{km}$. In this region, apart from an inner bubble
a few meters across, the Newtonian force is always much larger 
than the MOND force and also the MONDian acceleration 
$a_0\approx 10^{-10} \unit{m}\unit{s}^{-2}$. 
We recall the discussion in Section~\ref{theory},
where we noted that LPF would probe the regime $a_0<a_N<a_N^{trig}$,
with $a_N^{trig}\approx 10^{-5} \unit{m}\unit{s}^{-2}$. Indeed 
around $r_0\sim 400\unit{km}$, the Newtonian acceleration is
$a_N\sim a_N^{trig}$. 
The Newtonian tidal stress is therefore dominant in this regime
(with an intensity of the order $A\sim 10^{-11} \unit{s}^{-2}$),
but, crucially, it is approximately a DC 
component~\cite{Bekenstein:2006fi,Bevis:2009et}. This is to be contrasted
with the distinctively varying MONDian signal 
(see Fig.~\ref{fig:3bodyStress}) which, as we've shown, translates
into a signal peaking at frequencies where the noise is low. 
A DC component, on the other hand, is well buried in the $1/f$ noise. 
It is true that {\it in detail } 
the Newtonian tidal stress is not exactly constant
on the scale of $r_0$. But we do know what it is, to the same accuracy 
as we know the saddle location and trajectory, and can subtract it off. 
Furthermore its spectral shape away from its DC component is very different
from that predicted by the MOND signal, as shown in 
Fig.~\ref{fig:Tidal stresses ST1}. The Newtonian background
amounts to the subtraction of a known component.

\begin{figure}\begin{center}
\resizebox{1.\columnwidth}{!}{\includegraphics{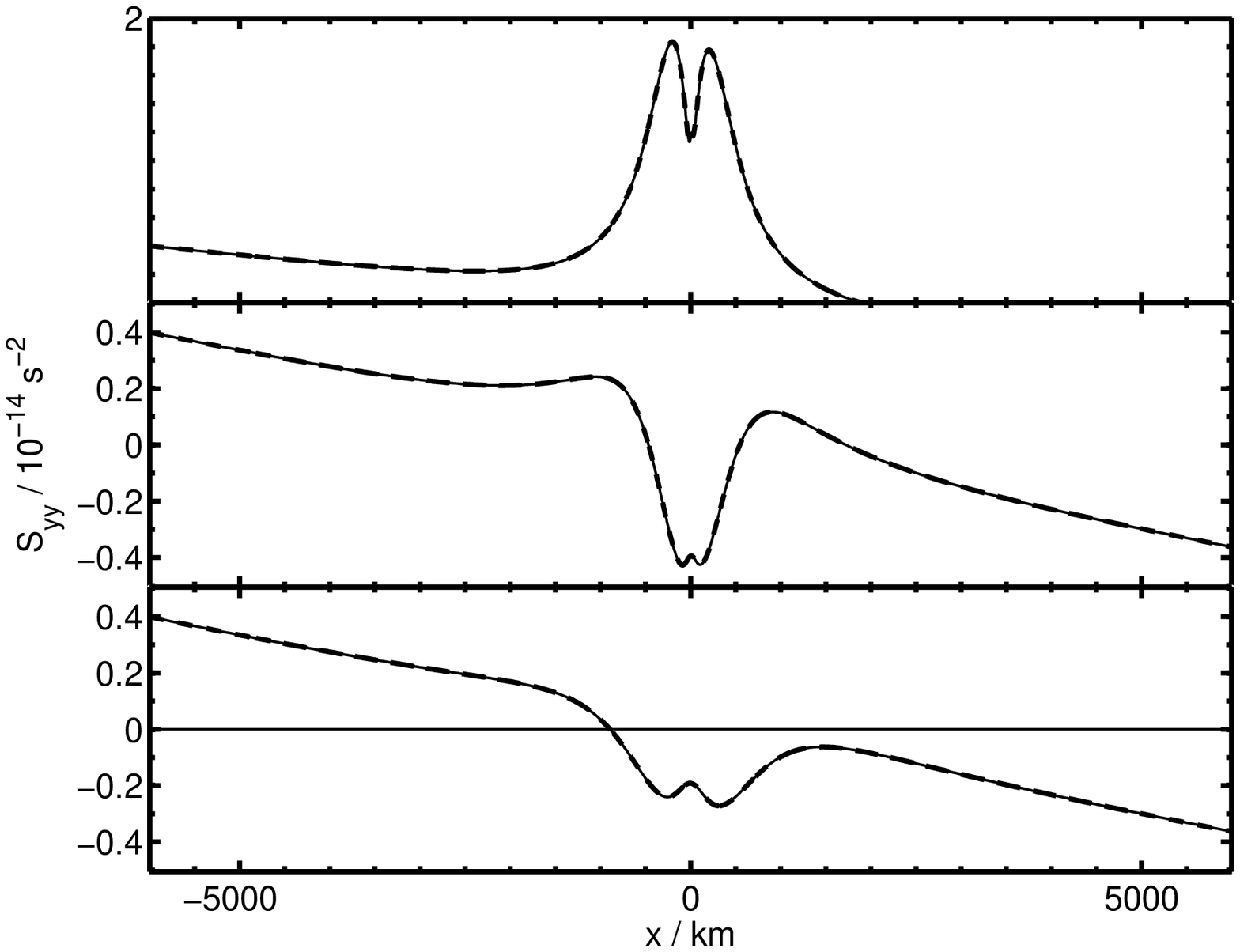}}
\resizebox{1.\columnwidth}{!}{\includegraphics{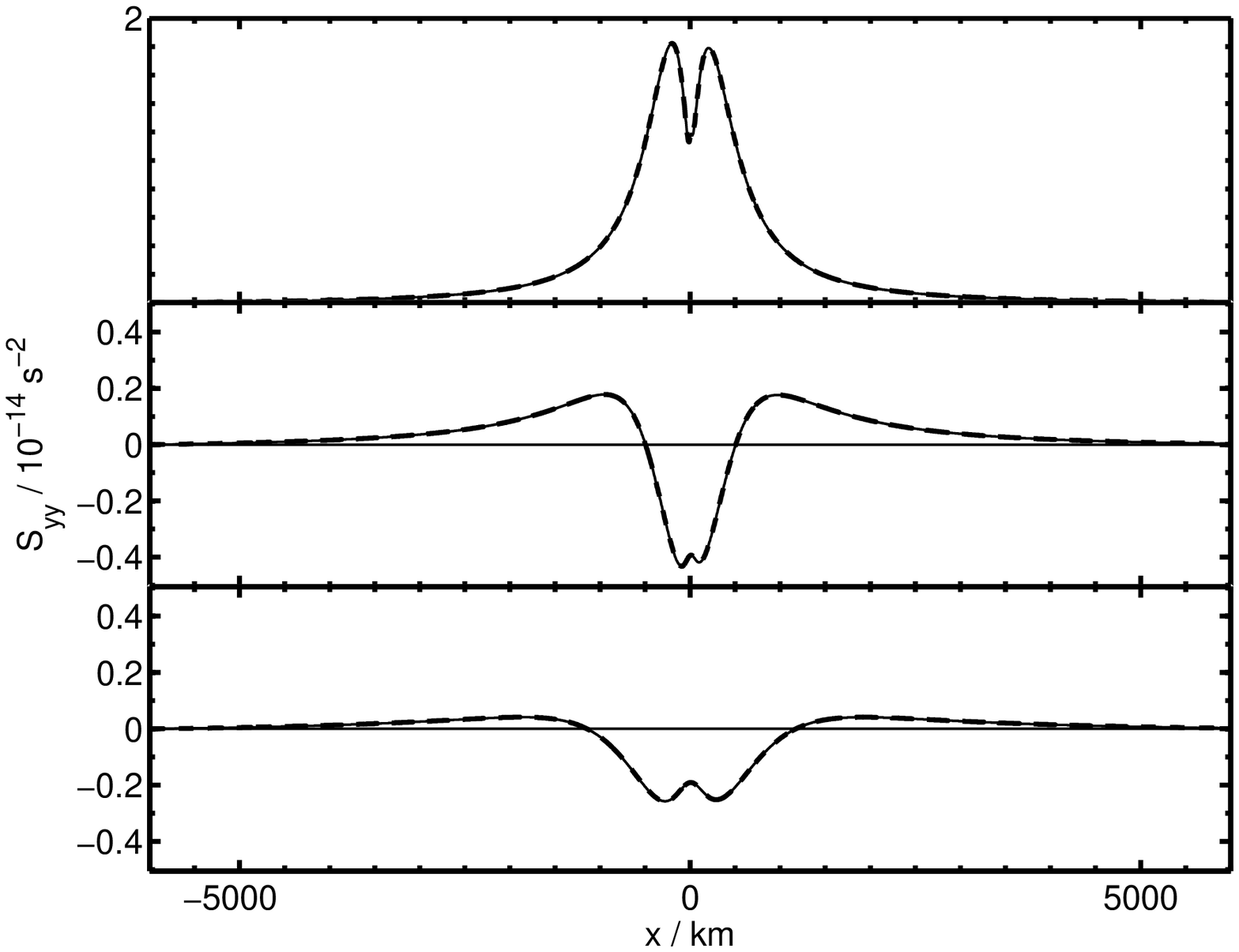}}
\caption{\label{fig:Noise ASD with Newt} This plot illustrates the 
systematic effects that might result from an incorrect 
Newtonian subtraction. We consider the transverse tidal stresses
felt in trajectories with impact parameters $b = 100, 500, 1000\unit{km}$.
We then subtract the DC constant Newtonian tidal stress 
contributions from $\phi$ (top) 
and its full contribution 
(bottom). As we can see an imperfect subtraction 
produces a spurious ramp in the stress.
} \end{center}
\end{figure}

A related matter was flagged in Section~\ref{secsnr}, and relates to 
subtracting off from $\phi$ its effect on the renormalization of the 
gravitational constant. Some of $\phi$ contributes to the Newtonian 
background and should not be included in the MONDian predictions
(cf. Eqn.~(\ref{Sij}); see also~\cite{Bevis:2009et}). 
In Figure~\ref{fig:Noise ASD with Newt}
we plotted the power spectrum of the non-DC component of the Newtonian
tidal stress produced by $\phi$. 
The impact of not subtracting the component of $\phi$ contributing to the
Newtonian measurement can be appreciated in 
Figure~\ref{fig:Noise ASD with Newt}.
This figure also gives us an idea of the
level of impact an imperfect Newtonian subtraction might have. 
We considered the transverse tidal stresses
felt in trajectories with impact parameters $b = 100, 500, 1000\unit{km}$.
In the top we subtracted only the DC component, in the bottom we subtracted
the full contribution of $\phi$ to the Newtonian tidal stresses. 
As we can see an imperfect subtraction 
would produce a spurious ramp in the stresses.

Another issue is LPF's self-gravity. The mission 
requirement is that the differential acceleration of the two LPF test
masses should be balanced at the level of 
$a\sim 10^{-9} \unit{m}\unit{s}^{-2}$, but actual performance
may beat the nominal requirement by a factor of 10. 
Yet again this is a DC component  and does not
affect the measurement in tidal stresses with the distinctive temporal
variations we have posited. There are of course time-varying 
uncertainties in the self-gravity balancing but these are much smaller.
They are mainly due to thermoelastic effects, and 
are on the level of $3\times 10^{-16}\unit{m}\unit{s}^{-2}/\sqrt{\unit{Hz}}$
at least down to 1mHz.

An issue related to this concerns the position of the saddle. Naively
one might think that with a self-gravity of the order of 
$10^{-9} \, \unit{m}\unit{s}^{-2}$ the position of the saddle would
be perturbed by the spacecraft.  The two test masses could even generate
distinct saddle points due to their gravity. This concern ignores
the fact that with realistic impact parameters we are {\it not}
testing the regime $a_N \sim a_0$, but the regime  $a_0<a_N<a_N^{trig}$
with much larger Newtonian accelerations. For instance 
for an impact parameter of 40~Km we have $a_N\sim 10^{-6}\unit{ms}^{-2}$. 
Around $b\sim r_0$ the Newtonian acceleration is
$a_N\sim a_N^{trig}\approx 10^{-5} \unit{m}\unit{s}^{-2}$.
We'd need to approach the saddle much closer than
about $400$ meters before self-gravity becomes an issue and the spacecraft
itself had to be included in the computation of the location of the saddle.
%For realistic $b$ one must not forget that we are probing a regime

%****** OLD to be reworked here
%It can be easily computed that if $\mu$ turns from 1 to a single
%power-law, then MONDian behavior in $\phi$ should be triggered
%when the Newtonian acceleration $a_N$ drops below $(4\pi/\kappa)^2
%a_0\sim 1.75 10^5 a_0\sim 10^{-5} m s^{-2}$. And indeed this is
%the rough Newtonian acceleration at ellipsoid $r_0$ in the 2-body
%problem worked out by Bekenstein and Magueijo.

\section{MONDian free-functions}\label{frefn}
In the second part of this paper we examine the generality of
our predictions. So far we have focused on type I theories
with a specific fitting function $\mu$ (the one 
used in~\cite{Bevis:2009et}).
But even if we restrict ourselves to type I theories 
there is a whole free function $\mu(z)$ to play with.
Would theorists be able to wriggle out of a negative
result availing themselves of this freedom? In the next 3 sections we
prove that under general conditions only type I theories 
with fine-tuned $\mu$-functions would survive 
a negative result (with notable exceptions to the rule, 
spelled out).
This conclusion is also expected to apply to type IIB theories,
although we won't prove it in detail. Type III and type IIA
theories turn out to be the only ones to evade a LPF saddle test. 

In this Section we start by reviewing previously proposed  $\mu$, 
laying down a common notation. We then discuss criteria for
physically permissible $\mu$, showing that for single power-laws 
they all produce SNRs of the same order, for impact parameters smaller
than 400~km. In Section~\ref{nullres} we will then consider
more structured functions $\mu$, with more regimes and scales.

\subsection{Notation and previous proposals}\label{notation}
As explained in Section~\ref{theory}, for 
type I theories two potentials act on non-relativistic test masses: the
Newtonian potential $\Phi_N$ and a fifth force $\phi$. Thus, the total
potential is $ \Phi=\Phi_N+\phi$,
or in terms of forces, ${\mathbf
F}={\mathbf F}_N+{\mathbf F}_\phi$ (as before, we 
set $\Xi=1$ for simplicity.). We recall that both
contributions satisfy Poisson type equations:
\be
\nabla^2\Phi_N=4\pi G \rho\label{PoissPhiN}
\ee
\be
\nabla\cdot (\mu \nabla \phi)=\kappa G\rho\label{Poissphi}
\ee
where, as before, we write $\mu$ with argument 
$z=\frac{\kappa}{4\pi}\frac{|\nabla\phi|}{a_0}$.
In the Newtonian limit $\mu\rightarrow 1$, whereas MONDian behaviour
in $\phi$ is triggered when $\mu\rightarrow z$.
Note that here
\be 
z={\sqrt\frac{y}{3}}
\ee 
where $y$ is the variable employed by Bekenstein in his original
paper on TeVeS~\cite{teves}. 
%We will provide a dictionary with other
%conventions, wherever needed for comparison with previous literature
Much confusion has arisen from different notations in this respect.

We should not confuse $\mu(z)$ with 
the function ${\tilde \mu}(x)$ used in 
type III theories (cf. Eqn.~(\ref{typeiii})) and also favoured
by astronomers. 
Even in type I (and also II) theories, we can loosely define 
an effective ${\tilde \mu}(x)$, obtained from adding equations 
(\ref{PoissPhiN}) and (\ref{Poissphi}) and comparing with 
Eq.~(\ref{typeiii}). This effective ${\tilde\mu}(x)$ function is
frequently used in fits to galactic phenomenology. However the two functions
${\tilde \mu}(x)$ and $\mu(z)$
can only be easily related if the MONDian curl term can be 
neglected. This proviso is often incorrectly ignored.
If the curl term is non-negligible, then type I theories don't properly 
have a $\tilde\mu(x)$ function, and there's no substitute for
integrating the equations on a case by case basis.

If the curl field can indeed be ignored in the integration of 
(\ref{Poissphi}), then it's easy to relate  
functions $\mu(z)$ and ${\tilde \mu}(x)$ (see, e.g.~\cite{kostasrev}). 
Using (\ref{PoissPhiN}) and (\ref{Poissphi}),
their definitions can then be rewritten as 
${\mathbf F}={\mathbf F}_N/{\tilde\mu}$ and ${\mathbf F}_\phi
=\frac{\kappa}{4\pi\mu}{\mathbf F}_N$, so that
${\mathbf F}={\mathbf F}_N+{\mathbf F}_\phi$ implies:
\be\label{mumu1}
{\tilde \mu}=\frac{1}{1+\frac{\kappa}{4\pi\mu}}\; .
\ee
In addition we can write the argument $x=F/a_0$ in terms
of $z=(\kappa/4\pi)F_\phi/a_0$ by deriving:
\be\label{mumu2}
x=\frac{4\pi}{\kappa}z\left(1+\frac{4\pi\mu(z)}{\kappa}\right)\; .
\ee
Eqns.~(\ref{mumu1}) and (\ref{mumu2})
provide a parametric expression for $\tilde\mu (x)$.
Note that Eq.~(\ref{mumu1}) trivially implies that
in the Newtonian regime ($\mu\approx 1$) MOND has the effect 
of renormalizing $G$ as:
\be\label{Gren}
G_{Ren}=\frac{G}{\tilde \mu}\approx G{\left(1+\frac{\kappa}{4\pi}\right)}\; .
\ee
a result already presented in Section~\ref{theory}.

Several $\mu$ functions have been previously proposed.
% (we have plotted
%a few in Figure~\ref{fig:Analytical mu}).
The  ``toy''
model used in Bekenstein's original paper~\cite{teves} 
follows from the implicit expression:
\be\label{bekmu1}
z^2=\frac{\mu^2(\mu-2)^2}{4(1-\mu)}\; .
\ee
%(In fact this is less of a toy than supposed improvements, as
%we shall see). 
A variation was employed  in~\cite{Bekenstein:2006fi} to 
facilitate analytical work on the 2-body problem:
\be\label{bekmu2}
z^2=\frac{\mu^2}{1-\mu^4}\; .
\ee
In some manipulations an inversion of the latter is useful:
\be
\mu={\sqrt\frac{\sqrt{1+4 z^4}-1}{2 z^2}}\; .
\ee
A proposal quite distinct from these two can be found in~\cite{Angus}:
\be\label{mualpha}
\mu(z)=\frac{z}{1-\frac{4\pi\alpha}{\kappa}z}
\ee
with the case $\alpha=1$ first suggested in~\cite{Zhao}. To bridge
our notation with the $\mu_s(s)$ used in~\cite{Angus} we should use
the dictionary (obtained from direct comparison of
(\ref{PoissPhiN}) and (\ref{Poissphi}) and their counterparts 
in~\cite{Angus}):
\bea 
\mu&=&\frac{\kappa}{4\pi}\mu_s\\
z&=&\frac{\kappa}{4\pi}s\; .
\eea
We stress that this function diverges, a property that sets it apart
from all those functions which tend to a constant as $z\rightarrow
\infty$. Underlying this statement is the postulate that the domain
of the function should be excised after the divergence is reached, i.e.
one should impose $s<1/\alpha$. The distinction between bounded and 
divergent $\mu$ was extensively studied in~\cite{FamaeyN} and is indeed 
central to the discussion\footnote{As a notational
word of caution note that for convergent free-functions we have
defined $\mu$  
such that $\mu\rightarrow 1$ as $z\rightarrow \infty$; 
whereas in~\cite{FamaeyN} one has 
$\mu_s\rightarrow \mu_0$, so that in effect we get the dictionary
$\mu_0=4\pi/\kappa$.}. A hybrid possibility, incorporating the behaviour
of (\ref{mualpha}) on galactic scales into a bounded function, can be adapted
from the proposal in~\cite{Zhao}, and will be examined 
in Section~\ref{nullres1}.

There is some debate over which ${\tilde \mu}$ functions best 
fit astrophysical data. Examples include~\cite{binney}:
\be
{\tilde \mu}(x)=\frac{x}{1+x}
\ee
and 
\be
{\tilde \mu}(x)=\frac{x}{\sqrt{1+x^2}}
\ee
or even~\cite{Zhao,Angus}:
\be\label{mutildealpha}
{\tilde \mu}(x)=\frac{2x}{1+(2-\alpha)x+\sqrt{(1-\alpha x)^2+4x}}\; .
\ee
With the proviso spelled out above (non-invertibility of a $\mu(z)$
in terms of a $\tilde\mu(x)$ in the presence of a curl field), function
(\ref{mutildealpha}) can be derived
from (\ref{mualpha}). Likewise (\ref{bekmu1}) and (\ref{bekmu2})
lead to:
\bea
{\tilde \mu}(x)&\approx&1+\frac{1-\sqrt{1+ 4x}}{2x}\nonumber\\
&=&\frac{\sqrt{1+4x}-1}
{\sqrt{1+4x}+1}\nonumber\\
&=&\frac{2x}{1+2x+\sqrt{1+4x}}
\label{bekmutilde}
\eea
where we have written three algebraically equivalent expressions to
facilitate comparison with the literature. Note that although (\ref{bekmutilde})
follows from (\ref{mutildealpha}) for $\alpha=0$, the same doesn't happen
with their $\mu$ functions, and Bekenstein's proposal (\ref{bekmu1})
is strictly not covered by (\ref{mualpha}). The claim has been 
made~\cite{Zhao,FamaeyN} that galactic observations
favour $\alpha=1$\footnote{As a further notational word of caution 
(c.f.~\cite{FamaeyN}) note that our $\tilde \mu$ tends to 
$G/G_{Ren}$ (or in the notation of~\cite{FamaeyN}, to $1/\nu_0$).
The Milgrom-like proposals considered above all tend to 1. Thus,
these proposals can only be approximately true with $G_{Ren}\approx G$.
Otherwise, the proposals considered here should strictly speaking be
labelled $\mu^{Milg}(x)$, with $\tilde\mu=\frac{G}{G_{Ren}}\mu^{Milg}$.}.

\subsection{Permissible, non-fine tuned $\mu$ functions}
Putting aside detailed predictions for galaxy rotation curves
(which may well have been combined with inconsistent approximations, e.g.
regarding the curl field), the following criteria are reasonable
for physically permissible, non-fine tuned $\mu$ functions
defining type I theories:
\begin{itemize}
\item A. The cosmologically measured $G$ cannot differ significantly from that
measured, say, by the Cavendish experiment. That is: $G_{ren}\approx G$.
\item B. When the total Newtonian acceleration $a_N$ drops below $a_0$
the full potential $\Phi$ must be in the MONDian regime, that is, we need
$\phi$ to be in the MONDian regime {\it and} to dominate $\Phi_N$.
\item C. Function $\mu$ should only have one scale, below which $\phi$ is
MONDian, and above which it is 
near Newtonian. The detailed form of the transition
is left undefined, but $\mu$ should have a single transition from $1$
to $z$.\end{itemize}
Items A and B have already been discussed in Section~\ref{theory}.
Item B is the most basic requirement for the theory to be of 
astrophysical use, {\it regardless of the details}. Item C has
been spelled out because it will be broken in the next Section, to
illustrate just how finely tuned $\mu$ would have to be
to evade a negative saddle result.

As explained in Section~\ref{theory}, these requirements
imply that $\phi$ must enter the MONDian regime at a much higher 
acceleration than $a_0$, leading to an intermediate regime 
$a_0<a_N<a_N^{trig}$ where $\phi$ is fully MONDian but still 
sub-dominant to $\Phi_N$. This implies that for 
{\it any} $\mu$ satisfying these constraints, 
when $a_N\sim a_0$ (i.e. for astrophysical applications)
we must necessarily have
\be\label{FphiM}
F_\phi\approx {\sqrt {F_N a_0}}\; .
\ee
This statement is independent of $\kappa$ and only relies on the fact
that $\mu\approx z=\frac{\kappa}{4\pi}\frac{|F_\phi|}{a_0}$
in the MONDian regime. {\it If
the curl term can be ignored} we therefore have $zF_\phi=\frac{\kappa}
{4\pi}F_N$, and thus (\ref{FphiM}) follows. Recalling $x=F/a_0$  we 
must conclude that:
\be
{\tilde \mu}(x)\approx\frac{F_N}{F}\approx 
1+\frac{1-\sqrt{1+ 4x}}{2x}\; .
\ee

The exception to this rule is obtained with a divergent $\mu$, as already 
announced in Section~\ref{theory}. 
Then, we discover the interesting behaviour~\cite{Angus,FamaeyN}
that $F_\phi$ goes to
a constant as $F_N$ grows to infinity, instead of becoming proportional
to $F_N$ (c.f. Eq.(\ref{rat})). Specifically, taking model
(\ref{mualpha}), we find that:
\be
F_\phi\approx \frac{a_0}{\alpha}
\ee
so that asymptotically no renormalization of $G$ takes place:
$G_{Ren}=G$. Such a functions would lead to a different $\tilde \mu$,
as we've seen. In Section~\ref{nullres1} we'll show that the MONDian 
behaviour driven by these functions would be rendered invisible to
LPF. 
%would be invisible to a LPF (see for example Fig.~\ref{fig:Fphi/FNnew}); 
%but it should be obvious that the argument for a large
%bubble breaks down, since $F_\phi/F_N$ never levels off. These functions
%do stand aside from those we are considering. We do not examine them
%in more detail because they contradict known Solar system constraints
%(see the discussion in ***).

\subsection{SNRs and $\mu$ dependence}
If we take the whole class of $\mu$ satisfying requirements I, II and III 
we conclude that they have the same $a_N^{trig}$ and consequently the same
$r_0$. Fiddling with $\mu$ therefore doesn't change the spatial scale
of the effect for type I theories (and
also for type IIB, but not type IIA or type III theories).
The predictions for $h(t)$ for $r<r_0$ are also model independent, since
they rely on $\mu\approx z$, for $z<1$. However the predictions
referring to regions with $r>r_0$ depend on the exact
form of the transient from $\mu\approx z$ to $\mu\approx 1$,
because they depend on $\delta\mu$, not $\mu$.
For example,
\be
\mu=\frac{z}{1+z}\approx 1-\frac{1}{z}
\ee
has a very different fall off from
\be
\mu=\frac{z}{\sqrt{1+z^2}}\approx 1-\frac{1}{2z^2}
\ee
a point recognized in~\cite{Bekenstein:2006fi}. Missing the saddle
by a lot more than $400\unit{km}$ would therefore leave us at the mercy
of model dependence, and $\mu$ functions satisfying A, B, C
could be found bypassing a negative result, e.g. 
\be
\mu=\frac{z}{(1+z^\beta)^{\frac{1}{\beta}}}\approx 1-\frac{1}{\beta z^\beta}
\ee
with a large $\beta$. However, for trajectories hitting the region $r<r_0$
(i.e. for $b<380\unit{km}$) the peak of the signal is actually 
model independent, and therefore the SNRs predicted aren't expected to
depend on the details of the theory.

\begin{figure}\begin{center}
\resizebox{1.\columnwidth}{!}
%{\includegraphics{SNRcontour.eps}}
{\includegraphics{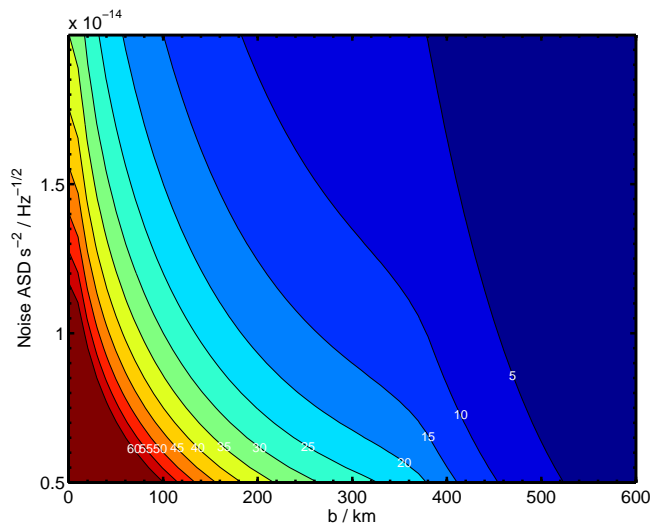}}
\caption{\label{fig:SNR contoursexcised}  Signal to Noise ratio
contours, for various impact parameters up to 600km and base noise
ASD, using the same templates as in Fig.~\ref{fig:SNR contours}  
but with an exponential fall off in 
the model-dependent region $r>r_0$. As we can see,
for impact parameters $b>400\unit{km}$ the SNR drops more sharply,
but nothing changes very much for $b<400\unit{km}$.
}\end{center}
\end{figure}

As an extreme illustration of the model (in)dependence of our
SNR predictions we have excised the signal outside the MOND bubble 
from our templates, imposing an exponential fall off. 
Fig.~\ref{fig:SNR contoursexcised} is the resulting
counterpart to Fig.~\ref{fig:SNR contours}.
We see that for $b<400\unit{km}$ our conclusions remain substantially 
the same. For $b>400\unit{km}$ the SNRs drop much 
more sharply. This is the worst that can be expected.

Impact parameters around $50\unit{km}$ are now considered within
easy reach. In order to bypass a negative result we would therefore 
have to shrink the bubble size, defined by scale $r_0$. This 
requires breaking condition C and consider contrived $\mu$ functions 
with two scales, which we now proceed to do in order to appreciate the
full implication of a negative result.

\section{A null result and designer $\mu$ functions}
\label{nullres}

It is often difficult to falsify a theory 
containing free parameters: all that can be readily done is to 
constrain its parameters. However the
constraints may be such that the theory becomes contrived beyond
``reasonable''. 
In what follows we imagine a scenario where no anomalies are found
with respect to the Newtonian expectation, for $b<400\unit{km}$.
Obviously all the theories considered so far
would be ruled out, to a degree of significance
of the same order as their expected SNR. The issue would then become
to determine which ``designer'' functions 
$\mu$ predict a SNR of order 1, 
%for a given impact parameter and noise ASD, 
thereby surviving a ``no anomaly'' result. 
The more contrived the required $\mu$, the more blatantly one should throw in
the towel.

In proposing a designer $\mu$ we shall impose that it satisfies requirements
A and B to the same extent as the functions we've been considering. 
%The theory should still be of astrophysical 
%use on very general grounds. The theory should not unduly renormalize
%$G$ and thus conflict with observations. No model dependent details
%go into these requirements: they are general common sense. 
The theory should still be of astrophysical 
use and not conflict with observations on very general grounds. 
However, we
drop requirement C, allowing the function to 
%be more contrived
%by permiting 
have two independent scales 
(notice that $a_N^{trig}$ is not independent for the models considered
so far). Specifically, we endow  $\mu$ with an 
intermediate power $n\neq 1$ linking the Newtonian regime, where 
$\mu\approx 1$, and the astrophysically relevant MONDian regime, 
where $\mu\approx z$.

\begin{figure} \begin{center}
\resizebox{1.\columnwidth}{!}{\includegraphics{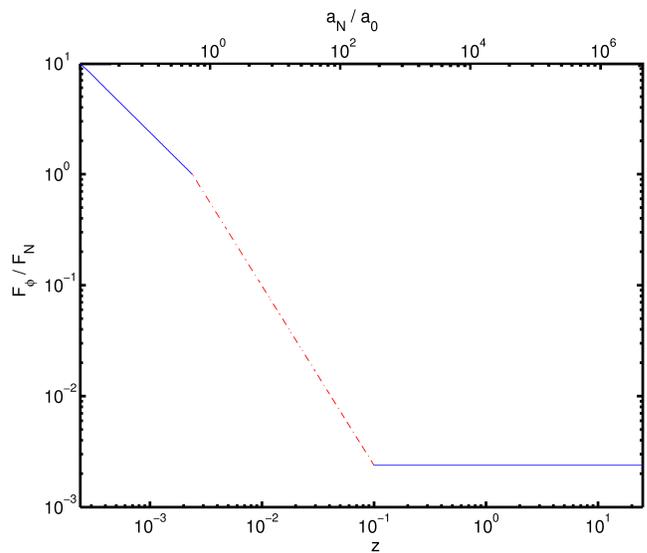}}
\caption{\label{fig:Fphi/FN plot null mu}{
Log plot of ratio
between the MONDian and Newtonian forces, $F_\phi/F_N$, against $z=(k
/ 4\pi) |F_\phi| / a_0$ (bottom axis) and $F_N/a_0$
(top axis).  So that $F_N\sim F_\phi$ when 
$F_\phi\sim a_0$ (and so $z=\kappa/4\pi$; also $F_N\sim a_0$) 
and at the same time
have $F_\phi/F_N\sim
\kappa  /4\pi\ll 1$ in the Newtonian regime ($z\gg 1$, $F_N\rightarrow\infty$),
we must trigger MONDian
behaviour in $\phi$ at accelerations much larger than $a_0$.
However, by allowing a sharper intermediate power-law in $\mu$,
the trigger acceleration $a_N^{trig}$ may be smaller (in this illustration
by a factor of 10).} 
}\end{center}
\end{figure}

Requirement B demands that $\mu\approx z$ for $z<\kappa/4\pi$, as
before, so that
$F_\phi\approx F_N$ when $F_N\approx a_0$, and 
$F_\phi\approx \sqrt{F_N a_0}$ for $a_N<a_0$. Requirement A imposes 
$\mu\rightarrow 1$ for large $z$, so that $G_{ren}$ is the same as for 
the single power-law $\mu$ considered before (cf. Eqn.~(\ref{Gren})). 
If we are to shrink the size of the MOND bubble so as to accommodate
a negative outcome from a saddle test, then we need 
a sharper power, $n>1$, bridging these two regimes.  Thus 
$F_\phi/F_N$ could increase faster, with decreasing $a_N$, 
from its small value $\kappa/4\pi$ in the Newtonian regime, to 1 at $a_N=a_0$. 
This would reduce $a_N^{trig}$ and thus $r_0$. The point is illustrated in 
Fig.~(\ref{fig:Fphi/FN plot null mu}), where we have replotted
Fig.~\ref{fig:Fphi/FN plot analytical mu} (made for a 
$\mu$ with a single power-law; 
recall the argument in Section~\ref{theory}).

\begin{figure} 
\begin{center}
\resizebox{1.\columnwidth}{!}
{\includegraphics{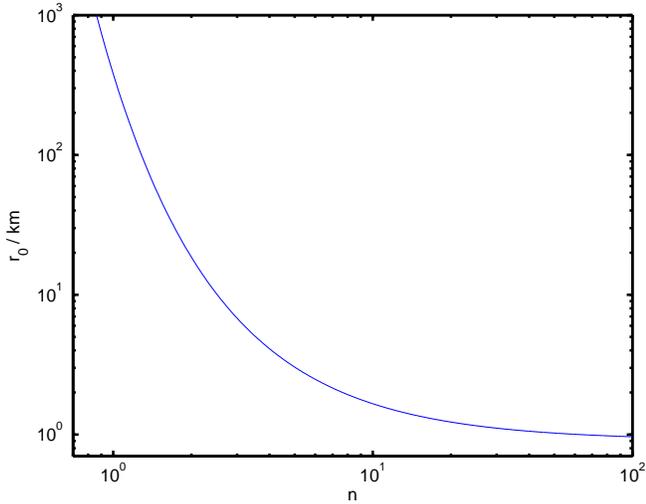}}
\caption{\label{r0n}{The size of the MOND bubble as a function of 
the intermediate power $n$. It is easy to collapse to bubble by
an order of magnitude (say to around $20\unit{km}$) with $n\sim 2$.
However, to make the bubble much smaller (say, on the order of a few 
kilometers), very dramatic intermediate powers would be required.} 
}
\end{center}
\end{figure}

These considerations fully specify the function $\mu$, up to 
details on the transition regions. We should have:
\bea
\mu\approx z&\quad {\rm for}\quad&
 z<\frac{\kappa}{4\pi}\\
\mu\approx \left(\frac{z}{z^{trig}}\right)^n
&\quad {\rm for}\quad &\frac{\kappa}{4\pi}<z<z^{trig}\\
\mu\approx 1&\quad {\rm for} \quad &z>z^{trig}
\eea
where the point where non-Newtonian behaviour in $\phi$ is triggered can be
interchangeably pinpointed by:
\bea
z^{trig}&=&\left(\frac{\kappa}{4\pi}\right)^{1-\frac{1}{n}}\\
a_\phi^{trig}&=&a_0\left(\frac{\kappa}{4\pi}\right)^{-\frac{1}{n}}\\
a_N^{trig}&=&a_0\left(\frac{\kappa}{4\pi}\right)^{-1-\frac{1}{n}}\; .
\eea
Notice that $a_N^{trig}$ is a now a truly independent parameter
of the theory (which can be traded for $n$). 
We still have that when $a_N<a_0$ the field $\phi$ dominates $\Phi_N$
as per requirement B, but now the intermediate region, where $\phi$
hasn't yet dominated but is already non-Newtonian, is in a narrower
band of accelerations $a_0<a_N<a_N^{trig}$. As a result the MOND
bubble shrinks according to 
\be
r_0\approx 380 \left(\frac{\kappa}{4\pi}\right)^{\frac{n-1}{n}}
\unit{km}\; .
\ee
In Figure~\ref{r0n} we have plotted this dependence. As can be seen, it's
easy to change $r_0$ by an order of magnitude with $n$ not much
different from 2. 
To reduce $r_0$ by more than that, however, a very extreme intermediate
power would be required\footnote{Notice that with 
this particular model the MOND bubble can never shrink 
smaller than $\frac{\kappa}{4\pi}380\unit{km}$.}. 
%($a_N^{trig}$ can never
%be lower than $\frac{\kappa}{4\pi}a_0$).

\begin{figure}\begin{center}
\resizebox{1.\columnwidth}{!}{\includegraphics
%{nullcontoursnr1.eps}}
{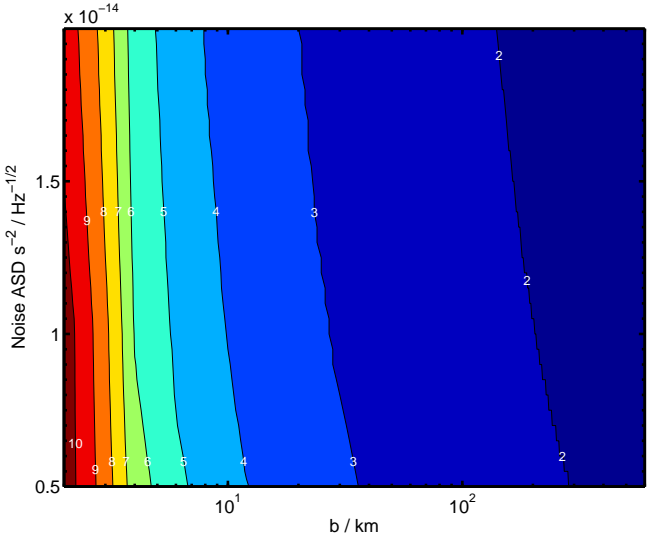}}
\caption{\label{fig:null SNR 1}{Contours of the power $n$ needed to
obtain SNR$=1$, for different noise levels and impact parameters
up to $b=600\unit{km}$. For 
$n\neq 1$ the function is ``unnatural''. We see that as soon as we
plunge deep into the MOND bubble, a rather unnatural designer
$\mu$ becomes necessary to accommodate a negative result.
}
}\end{center}
\end{figure}

Regrettably we can never make a model independent statement on what
$n$ is needed for a SNR of order 1. If nothing is observed then 
by the nature of the problem we must be making observations in the regime
$b\gg r_0(n)$. Therefore we are necessarily probing the transient
from $\mu\propto z^n$ to $\mu\sim 1$, dependent on the exact form 
of the function $\mu$. Nonetheless it is interesting to perform
this exercise, assuming a specific function, say:
\be \mu(z) = \frac{\left(\frac{z}{z^{trig}}\right)^n}{1
+ \left(\frac{z}{z^{trig}}\right)^n} \; .\ee 
For $z\gg z^{trig}$ this can be expanded as:
\be \mu 
\approx 1 + \delta\mu= 1 -\left(\frac{z^{trig}}{z}\right)^n\; .
\ee  
Also for $b\gg r_0(n)$ 
the curl field is negligible, so we can write:
\be
\mu \vF_\phi = \frac{\kappa}{4 \pi} \vF_N \label{mu grad phi
sym}\ee  
and solve it perturbatively. Expanding 
as in $\vF_\phi={}^0\vF_\phi+\delta \vF_\phi$, we have
to zero order ${}^0\vF_\phi=\frac{\kappa}{4\pi}\vF_N$. To first order
we then obtain:
\be 
\delta \vF_\phi \approx - \frac{\kappa}{4 \pi} (\delta\mu)\vF_N
\approx 
\left(\frac{4 \pi}{\kappa}
\frac{a_0}{|\vF_N|}\right)^n\vF_N \label{order of mag
null} \ee  from which the tidal stresses can be inferred. 

The results are condensed in Fig.~\ref{fig:null SNR 1}, depicting
the value of $n$ needed for a given $b$ and noise level in order for
a SNR of one to be obtained (and so a negative result be acceptable). 
As we can see as soon as we
plunge deep into the MOND bubble, a rather unnatural designer
$\mu$ becomes necessary to accommodate a negative result.

\subsection{Motivated functions with features}\label{nullres1}

In this paper we have attempted not to mix galaxy rotation fits with
our considerations. The reason is that it's not clear how these fits
would stand if performed together with the need to fit Solar system data
and a saddle test: the performance of goodness of fit statistics under
{\it joint} constraints might be very different. Also 
penalization for extra parameters (such as $\alpha$)
has probably not been properly
enforced (see~\cite{sorkin} for an example of the implications), 
and it would certainly
behave very differently in a joint fit, where the number of degrees of 
freedom would be much larger (c.f. the behaviour of the 
Bayesian information criterion as seen in~\cite{sorkin}). 
Having said this, it may
well be that what we labeled  ``two-scale'' or ``designer''
models is precisely 
what is needed for one such joint fit. We therefore examine
how these motivated functions fare in terms of SNR for a saddle test.

\begin{figure} \begin{center}
\resizebox{1.\columnwidth}{!}{\includegraphics{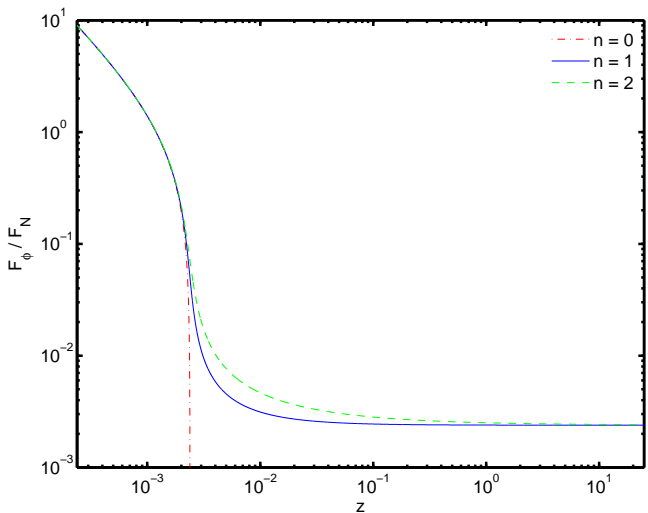}}
\caption{\label{fig:Fphi/FNnew}
{Log-log plot of ratio
between the MONDian and Newtonian forces, $F_\phi/F_N$, against $z=(k
/ 4\pi) |F_\phi| / a_0$ for functions 
(\ref{designmot}), with $n=0,1,2$. For $n=0$ we realize the 
divergent function (\ref{mualpha}) and as we can see there are
only two regimes, corresponding to two power-laws, with the ratio
never flattening to a constant. For all other cases we have a
3 piece function, with a fall off to the Newtonian regime which depends
crucially on $n$.} 
}\end{center}
\end{figure}

One type of function which evades a saddle test are those  
$\mu$ which diverge (eg.~\cite{Angus}). As explained before,
then the asymptotic $G$ is not renormalized, since $F_\phi$ tends to 
a constant, 
\be
{\mathbf F}_\phi\approx \frac{a_0}{\alpha}\frac{{\mathbf F}_N}
{F_N}\; ,
\ee 
as $F_N\rightarrow\infty$. Thus the profile of $F_N/F_\phi$ is actually 
not ``designer'' in the sense we have been using the word, since it merely
changes from one power-law ($1/\sqrt{F_N}$) to another ($1/F_N$),
never leveling off into a constant. The counterpart of 
Fig.~\ref{fig:Fphi/FN plot analytical mu} for this model is plotted
in Fig.~\ref{fig:Fphi/FNnew}, curve labelled $n=0$. Since $F_\phi/F_N$
never levels off, $G_{Ren}=G$, with $a^{trig}\approx a_0$. A detailed
analysis of the saddle effects of this theory cannot be carried out
with the methods proposed in this Section. Since $\mu$ doesn't go to
a constant, strictly speaking full MOND-like effects are present for
all accelerations, and the curl field can never be neglected. Still,
one may expect the order of magnitude of the predicted effects to be small,
and the associated saddle bubble to be invisible for LPF. We defer to a future
publication a more detailed examination of the predictions associated
with these functions.

\begin{figure}\begin{center}
\resizebox{1.\columnwidth}{!}{\includegraphics
%{nullcontoursnr1.eps}}
%{snrnullcontoursv3.eps}}
{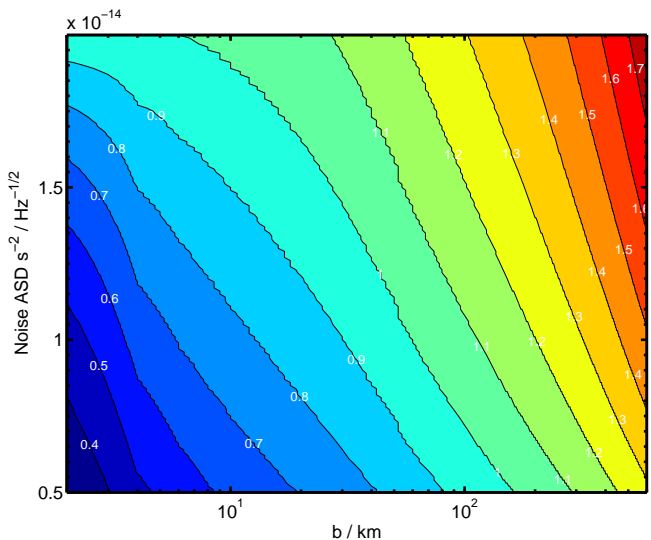}}
\caption{\label{fig:null SNR 2}{Contours of the power $n$ 
for models (\ref{designmot}) required in order to 
obtain SNR$=1$, for different noise levels and impact parameters
up to $b=600\unit{km}$. These are to be seen as an upper bound on 
parameter $n$, should a negative result from a LPF saddle  experiment
be found, for a given experimental set up.
}
}\end{center}
\end{figure}

Divergent $\mu$ functions, however, 
{\it may} fall foul of Solar system 
constraints (see discussions in~\cite{Fam-gaugh,FamaeyN,ssconst,Gentile},
but we stress that this conclusion might not be fully general).
A compromise can be struck by combining the functional
form of the proposed unbounded $\mu$ with a curve flattening to a constant
beyond the scales probed by galaxy rotation curves. With such constructions
we are plainly entering the realm of the 
``designer'' (or multiple functional form) functions, 
as defined in this Section. Such multiple regime, multi-scale, functions
are implemented in one way or another, in all such 
proposals~\cite{Zhao,Bruneton,McGaughMW,Zhaoeco}.
For example, translating~\cite{Zhao} into our variables
leads to:
\be\label{designmot}
\frac{4\pi z}{\kappa}=\frac{\mu}{\frac{\kappa}{4\pi}+\alpha \mu}
\frac{1}{(1-\mu)^n}\; .
\ee
On galactic scales this reduces to (\ref{mualpha}), yet for the purpose
of the Solar system we have:
\be
\mu\approx 1-{\left(\frac{a_0}{\alpha F_\phi}\right)}^{\frac{1}{n}}\;.
\ee
The $F_\phi/F_N$ profile for these models is plotted in 
Fig.~\ref{fig:Fphi/FNnew}, which should be compared to the ``natural''
Fig.~\ref{fig:Fphi/FN plot analytical mu} and the more contrived toy
model depicted in Fig.~\ref{fig:Fphi/FN plot null mu}. These models
can be constrained using the methods proposed in this Section.
In the face of a negative result, as above, we'd be able to  
assume we are in the 
quasi-Newtonian regime, to conclude, after a similar argument,
the counterpart of (\ref{order of mag null}):
\be \label{deltaF1}
\delta \vF_\phi \approx - \frac{\kappa}{4 \pi} (\delta\mu)\vF_N
\approx \frac{\kappa}{4\pi}
\left(\frac{4 \pi}{\alpha \kappa}
\frac{a_0}{|\vF_N|}\right)^{\frac{1}{n}}\vF_N  \ee
We can now constrain parameter $n$ as before, with the result plotted
in Fig.~\ref{fig:null SNR 2}. As in Fig.~\ref{fig:null SNR 1} we have
plotted the value of $n$ (not to be confused with the parameter used there)
for which the SNR turns out 1, for a given noise and impact parameter.
A negative result from LPF would therefore require $n$ to be smaller
than this value, which is therefore an upper bound.

\section{Conclusions}\label{concs}
In this paper we showed how a LPF saddle flyby would either
detect MOND to a high SNR or rule it out, if not comprehensively,
at least to a large extent.
The former conclusion could be expected.
Even though in Sections~\ref{secsnr} and~\ref{systematics} we 
provided rigorous and 
quantitative SNR estimates, the high levels forecast can be
understood with a ``back of the envelope'' calculation.
The exercise highlights an uncanny coincidence.
The accelerometer aboard LPF has a non-white noise profile, dipping
in the region of the mHz, i.e. in the rough time scale of minutes. The
motivation for this design lies in 
the gravitational wave signals to be targeted by LISA. It just so
happens that the MONDian
bubbles of anomalous tidal stresses around the Earth-Sun-Moon
saddles have a length scale of the order of a hundred kilometers. 
Anything free-falling in the Earth-Moon region 
has a typical speed of the order of 1km/s. Thus, 
the time scale for crossing a MONDian bubble will be of the order of minutes:
just where the instrument performance is optimal. This is
a remarkable coincidence. Scribbling on the
back of an envelope, using the expression for the SNR of a noise-matched
filter and the order of magnitude of the stresses and noise, 
promptly reveals double figure SNRs.

The question then arises as to how generic this conclusion is, or conversely,
should a negative result be found, how thoroughly have we ruled out
MOND. In Section~\ref{theory} we scanned the full array of MONDian 
theories, showing that in the non-relativistic regime they fall into
only 3 categories  (which we labelled type I, IIA/B and III). Even if 
we restrict ourselves to type I theories (a class containing the vast 
majority of these theories, including TeVeS), we benefit from the
leverage of a free function, $\mu$. 
Could MONDologists use $\mu$ to survive a negative result?

In Sections~\ref{frefn} we examined $\mu$ functions on offer in the
literature and laid down criteria for reasonable $\mu$ based on
astrophysical usefulness, viability in the face of constraints, 
{\it and naturalness}. We found that once these criteria are taken into account
the size of the MOND bubble, $r_0$, is fixed. There are exceptions to
this rule; e.g. diverging $\mu$ functions, but these {\it may} be 
subject to other constraints (they will be examined elsewhere).
Predictions for what 
happens inside the bubble are also model independent; however the
tidal stress anomalies outside the bubble depend on the transient
from MONDian into Newtonian regime, with a fall-off which is indeed model
dependent. Thus, for impact parameters smaller than
$r_0$ the predicted SNRs are robust, and do not change substantially with 
the model. For the currently expected $b$ (around $50\unit{km}$, 
with $r_0\sim 380\unit{km}$) this is indeed the case.

Therefore the only way for MOND to wriggle out of a negative LPF result 
would be to change the bubble size $r_0$. This can only be accomplished 
with ``designer'' $\mu$-functions. If $\mu$ is allowed to
have two scales and two power-laws away from its Newtonian value of 1, then
it is possible to bypass a negative LPF result. Even for 
undemanding noise levels and impact parameters, the intermediate
power becomes very contrived. Similarly fine-tuned functions have been
proposed in the literature. In Section~\ref{nullres1}, we showed how
LPF could be used to constrain them. The point remains that one
would have to bend backwards to accommodate a negative result,
within type I theories. There are exceptions to this rule (for example
diverging $\mu$ functions) and these derserve further attention.

Although we didn't present quantitative results for type II
theories, the same conclusions apply qualitatively 
to type IIB theories (but not type IIA theories). 
If $G$ is renormalized and the free function $\nu$ is
chosen to produce the same phenomenology as type I theories
(in particular with regards to $G_{Ren}$ and MONDian behaviour), then
the MOND bubble has the same size, and the anomalous tidal stresses are
of the same order.  As explained in Section~\ref{theory}, 
in both types of theory MONDian behaviour is due
to an extra field $\phi$, and if one attends simultaneously to 
$G_{Ren}\approx G$ and $\phi\sim \Phi_N$ for $a_N\approx a_0$,
then MONDian behavior in $\phi$ should be triggered at the same Newtonian 
acceleration $a_N=a_N^{trig}\gg a_0$. This implies a MONDian bubble of the 
same size $r_0$. Furthermore the (also $\nu$-independent) effects 
inside the bubble are different from type I predictions, but 
stronger. Type II theories don't have a curl field (in the sense
define above), a feature which 
softens the anomalous tidal stresses in type I 
theories~\cite{Bekenstein:2006fi}. 
A detailed quantitative prediction for type II theories 
is currently being investigated (see~\cite{aliqmond}; also we reference
here a paper on the matter which has appeared since the present paper was 
submitted~\cite{typeII}.)

However, it may be that the relativistic
mother theory is set up in such a way that the cosmological and 
non-relativistic $G$ coincide, the case of type IIA theories. 
In this case the MOND bubble around the saddle is very small. Likewise type 
III theories (or, rather, its single relativistic
realization, the Einstein-Aether theory) produce effects around
saddles which are unobservable with current technology. 
In these theories $G$ is not renormalized and
$a^{trig}=a_0$, so that the MOND bubble is a few meters across. 
Remarkably, solar system tests are extremely constraining 
upon type III theories, due to the so-called external field 
effect~\cite{Blanchet}.  By contrast 
solar system effects for type I and IIB theories are suppressed by
a factor of $\kappa/4\pi$. Thus saddle tests and planetary 
orbits seem to be complementary in constraining MONDian theories.

We close by noting that we could, of course, detach our considerations 
entirely from the MOND paradigm (as an alternative to dark matter),
and regard these theories formally as a class on alternative theories 
of gravity (see~\cite{Clifton11} for an extensive review). It is 
remarkable that only three classes of theories emerge in the 
non-relativistic regime, which we labelled type I, II and III in 
Section~\ref{theory}. We could then view $\kappa$ and $a_0$ as free 
parameters, converting a LPF saddle flyby into a constraint or a detection 
in this space. We are currently working on this alternative approach.

\begin{acknowledgments}
We'd like to thank Luc Blanchet, Tim Clifton, Benoit Famaey, Pedro Ferreira,
Martin Hewitson, Natalia Korsakova, Mordehai Milgrom, 
B.S. Sathyaprakash, Christian
Trenkle and an anonymous referee for comments and suggestions. We're also
grateful for the input from the whole LPF science team, provided at 
two meetings at RAL and IC. Our numerical work 
was performed on the COSMOS supercomputer, which is
supported by STFC, HEFCE and SGI. 
\end{acknowledgments}

\bibliography{references}

\end{document}